%% file: main-AliEdge.tex
\documentclass[sigconf]{acmart}

\usepackage{tikz}
\usepackage{amsmath}

\usepackage{filecontents}

\makeatother
\usepackage{hyperref}
\usepackage{multirow}

\usepackage[utf8]{inputenc}
\usepackage[ruled, lined, longend, linesnumbered]{algorithm2e}
\usepackage{fontawesome}

\input{header}

\usepackage{array}
\newcolumntype{L}[1]{>{\raggedright\let\newline\\\arraybackslash\hspace{0pt}}m{#1}}
\newcolumntype{C}[1]{>{\centering\let\newline\\\arraybackslash\hspace{0pt}}m{#1}}
\newcolumntype{R}[1]{>{\raggedleft\let\newline\\\arraybackslash\hspace{0pt}}m{#1}}

\newcommand{\revise}[1]{{{#1}}}

\newcommand{\sys}{NEP\xspace}
\newcommand{\node}{site\xspace}
\newcommand{\nodes}{sites\xspace}
\newcommand{\cloudsys}{AliCloud\xspace}
\newcommand{\implications}[1]{
	\textbf{\underline{Implications}}
	\textit{#1}}

\copyrightyear{2021}
\acmYear{2021}
\setcopyright{acmcopyright}\acmConference[IMC '21]{ACM Internet Measurement
	Conference}{November 2--4, 2021}{Virtual Event, USA}
\acmBooktitle{ACM Internet Measurement Conference (IMC '21), November 2--4,
	2021, Virtual Event, USA}
\acmPrice{15.00}
\acmDOI{10.1145/3487552.3487815}
\acmISBN{978-1-4503-9129-0/21/11}

\begin{document}

	\title{From Cloud to Edge: A First Look at Public Edge Platforms}

	\author{Mengwei Xu$^1$, Zhe Fu$^2$, Xiao Ma$^1$, Li Zhang$^1$, Yanan Li$^1$, Feng Qian$^3$, Shangguang Wang$^1$, Ke Li$^4$, Jingyu Yang$^4$, Xuanzhe Liu$^5$}
	\affiliation{%
	\institution{Beijing University of Posts and Telecommunications$^1$}
		\country{}
	}
	\affiliation{%
		\institution{Tsinghua University$^2$, University of Minnesota - Twin Cities$^3$}
		\country{}
	}
	\affiliation{%
		\institution{Unaffiliated$^4$, Peking University$^5$}
		\country{}
	}
	\renewcommand{\shortauthors}{Mengwei Xu et al.}

	\begin{abstract}
		\input{abs}

	\end{abstract}



	\maketitle


	\input{todo}
	\input{intro_new}
	\input{sys}
	\input{performance}
	\input{analysis}
	\input{discuss}

	\input{related}
	\input{conclusion}
	
	\section*{Acknowledgments}
	Mengwei Xu was supported by National Key R\&D Program of China under grant number 2020YFB1805500, the Fundamental Research Funds for the Central Universities, and National Natural Science Foundation of China under grant number 61922017.
	Xuanzhe Liu was supported in part by Alibaba University Joint Research Program.
	We hereby give special thanks to Alibaba Group for their contribution to this paper.
	We also thank our shepherd, Aaron Schulman, and the anonymous IMC reviewers for their useful suggestions.
	Shangguang Wang is the corresponding author of this work.

	\bibliographystyle{plain}
	\bibliography{ref}

	\input{appendix-billing}

\end{document}

%% file: header.tex
\usepackage{wrapfig}
\usepackage{comment}
\usepackage{colortbl}
\usepackage{tabularx}

\usepackage{color,soul}
\usepackage{graphics}
\usepackage{graphicx}
\usepackage{lipsum}

\usepackage{CJKutf8} 

\usepackage{tikz}
\usetikzlibrary{decorations.pathmorphing, patterns,shapes,snakes} 

\usepackage[inline]{enumitem}	
\usepackage{listings} 
\usepackage{lipsum}

\usepackage{capt-of}	

\usepackage{ulem}  

\ifdefined\HCode
\usepackage[compatibility=false]{caption}

\else
\usepackage{subcaption} 
\fi

\usepackage{mathrsfs}	
\usepackage{amsfonts}

\usepackage[export]{adjustbox} 

\definecolor{myForestGreen}{RGB}{34, 139, 34}
\definecolor{airforceblue}{rgb}{0.36, 0.54, 0.66}

\renewcommand{\paragraph}[1]{\vskip 3pt\noindent\textbf{#1 }}	 

  {\begin{list}{$\bullet$}%
     {\setlength{\parsep}{0pt}%
      \setlength{\topsep}{0pt}%
      \setlength{\itemsep}{2pt}}}%
  {\end{list}}
\newcommand\Note[1]{\sethlcolor{green} \hl{#1}} 

\newenvironment{myitemize}%
  {\begin{itemize}
	[leftmargin=0cm,
		itemindent=.3cm,
		labelwidth=\itemindent,
		labelsep=0pt,
		parsep=3pt,
		topsep=2pt,
		itemsep=1pt,
		align=left]
  }%
  {\end{itemize}}    

  {\begin{enumerate}
	[leftmargin=0cm,itemindent=.5cm,labelwidth=\itemindent,
		labelsep=0pt,
		parsep=1pt,
		topsep=1pt,
		itemsep=3pt,
		align=left]
  }%
  {\end{enumerate}}    
  
  {\begin{enumerate*}
	[label=\roman*)]
  }%
  {\end{enumerate*}}      

  {\begin{enumerate*}
	[label=\roman*)]
  }%
  {\end{enumerate*}}

\input{terms}
\makeatletter
\def\@copyrightspace{\relax}
\makeatother

%% file: abs.tex
Public edge platforms have drawn increasing attention from both academia and industry.
In this study, we perform a first-of-its-kind measurement study on a leading public edge platform that has been densely deployed in China.
Based on this measurement, we quantitatively answer two \textit{critical} yet \textit{unexplored} questions.
First, from end users' perspective, what is the performance of commodity edge platforms compared to cloud, in terms of the end-to-end network delay, throughput, and the application QoE.
Second, from the edge service provider's perspective, how are the edge workloads different from cloud, in terms of their VM subscription, monetary cost, and resource usage.
Our study quantitatively reveals the status quo of today's public edge platforms, and provides crucial insights towards developing and operating future edge services.

%% file: todo.tex

%
%
%
%
%
%
%
%


%% file: intro_new.tex
\section{Introduction}
By bringing computation and storage closer to end users, edge computing is expected to benefit a wide range of applications such as auto-driving, AR/VR, IoTs, and smart cities.
Edge computing can be instantiated by various paradigms such as cloudlet~\cite{satyanarayanan2009case} and MEC~\cite{mec-whitepaper}.
This work targets at \textit{public edge platforms} (or edge clouds), which deploy massive yet lightweight datacenters (DCs) decentralized at different geographical locations and provide hardware resources to third-party customers.
Such platforms are increasingly popular,
because they inherit the key spirits from commercial cloud computing that has proved its tremendous success in the past decade.
For example, major cloud providers are building their public edge platforms such as Azure Edge Zone~\cite{azure-edge-zone} and AWS Local Zones~\cite{aws-local-zones}.

Edge platforms offer several major advantages compared to classic cloud computing, such as much lower network latency, improved application performance, and potentially reduced operational costs.
Although these benefits are qualitatively known, their \textit{quantitative characteristics} in operational environments
are far from being comprehensively studied.
In this paper, we conduct to our knowledge the first measurement study of a commercial public edge platform in the wild,
from both the end users' and the edge providers' perspectives. For the former, we investigate key metrics that
are perceivable by edge customers (who deploy their apps on edges), such as end-to-end latency, throughput, and application QoE; for the latter, we take a closer look at the edge workload dynamics.
Such a ``dual'' approach helps reveal a complete landscape of the edge ecosystem.

\input{tab-density}

\textbf{Challenges} We face several challenges in this study.


First, edge servers are more geographically distributed compared to traditional cloud servers (Table~\ref{tab:density}), thus requiring more effort on conducting measurements at a large number of vantage points.
To this end, we perform a country-wide crowd-sourced study involving 158 participants, who run our custom testing tool on their mobile devices.
We obtained meaningful results from 41 cities in China over diverse access networks (WiFi/LTE/5G). We also develop two user applications that can benefit from edge computing:
cloud gaming and live video streaming, and deploy them over commercial edge/cloud services. Our own implementations allow us to instrument the apps and obtain detailed QoE information.

Second, obtaining an insider's view of operational edge service providers is difficult.
In this study, we collaborate with a commercial, multi-tenant edge service provider, referred to as
\sys (\textbf{N}ext-generation \textbf{E}dge \textbf{P}latform\footnote{\revise{\sys is commercially known as Alibaba ENS~\cite{ens}.}}).
As a leading edge service provider in China, \sys has operated for more than 3 years, serving a wide spectrum of applications used by millions of users.
The deployment scale of \sys is significantly larger than the aforementioned edge platforms or popular cloud platforms as summarized in Table~\ref{tab:density}.
We collected detailed usage traces of \textit{all}
Infrastructure-as-a-Service (IaaS) VMs in \sys's DCs for three months, and use them to profile the edge workloads.

Third, ideally we would like to quantitatively compare edge to cloud in terms of their performance and server workload.
Through active measurements, we compare \sys's performance with Alibaba Cloud ECS (\cloudsys)~\cite{ali-ecs}, a leading cloud provider in China.
The server workload comparison is much more challenging as few cloud providers release their workload traces.
To this end, we compare our \sys dataset with the Azure cloud dataset~\cite{cortez2017resource} collected in 2019 -- the only touchable, full cloud workload that we are aware of~\footnote{$\S$\ref{app:traces} summarizes the publicly available workload traces and clarifies the reason why Azure dataset is the only appropriate one for comparison.}.
\revise{We admit that this comparison is not perfectly apples-to-apples as the Azure data is mainly for the U.S. market and the data collection period is different.
Yet, this is the best we can achieve due to a lack of publicly available workload traces.
}
We therefore draw our conclusions in a conservative and cautious manner -- only when we are confident that the disparities we show between the \sys and Azure datasets are very likely attributed to the inherent differences between cloud and edge.


\textbf{Findings} We summarize our key findings below.

\textit{(1) \textbf{Network latency} ($\S$\ref{sec:network-delay})} is the key metric that edges are expected to improve.
Our crowd-sourced results show that \sys offers a lower network delay: the median RTTs between users and the nearest edge DC are \revise{10.5ms/34.2ms/11.7ms} for WiFi/LTE/5G networks, which are \revise{1.89$\times$/1.42$\times$/1.35$\times$} lower than the nearest cloud DC of \cloudsys, respectively.
The reduced network jitter is even more significant ($\sim$5$\times$).
However, \sys still cannot (or barely)
meet the requirements of delay-critical applications like cloud VR/AR (5ms--20ms)~\cite{cloud-vr} and auto-driving (10ms)~\cite{v2x-whitepaper}.
This is because the nearest \revise{server of \sys} is still 5--12 (median: 8) hops away from end users, instead of 1--2 hops as commonly envisioned for edge computing~\cite{mec-whitepaper,satyanarayanan2009case}.
The \revise{network performance of \sys} can therefore be further improved by a denser deployment of DCs and by sinking DCs into the ISP's core networks or even cellular base stations.

\textit{(2) \textbf{Network throughput} ($\S$\ref{sec:throughput})}
We find that \revise{by bringing servers closer to users, \sys improves network throughput} only when the last-mile bandwidth capacity is high enough, e.g., $>$200Mbps like 5G downlink.
Otherwise, e.g., for WiFi and LTE, the end-to-end network throughput is bottlenecked by the wireless hop instead of the Internet;
therefore edges exhibit no improvements over remote clouds.
Considering the rare use cases where such high throughput is demanded by today's application and its incurred high operational cost, we believe that throughput is not a primary advantage of \revise{\sys-like edges} at this moment.
However, this situation may change in the near future, when 5G shifts the bottleneck from the last mile to the wired Internet.

\textit{(3) \textbf{Application QoE} ($\S$\ref{sec:app-performance})}
%
Through controlled experiments, we observe that
placing the gaming backend on nearby \revise{\sys \nodes} can noticeably improve the response delay compared to remote clouds (91ms vs. 145ms).
To further enhance the QoE, optimizations shall focus on server-side gaming execution, e.g., through higher CPU parallelism or hardware acceleration.
For live streaming, \sys only brings modest improvement (up to 24\% of streaming latency) and the streaming delay remains high (400ms without a jitter buffer).
The reasons are twofold: (i) \revise{\sys's edge resources} can effectively reduce the propagation delay, but not necessarily the transmission delay;
(ii) more importantly, the bottleneck is oftentimes the content processing/computation rather than the network.
Future efforts should thus focus on improving the hardware capacities (e.g., the camera's image signal processor) and the system-software stacks.

\textit{(4) \textbf{Characteristics of \revise{\sys's edge VMs}} ($\S$\ref{sec:vm-type})}
We find that \sys's VMs and their incurred workloads are noticeably different from those of Azure.
%
\sys's VMs are often equipped with more resources, including both CPU (median: 8 vs. 1) and memory (median: 32GBs vs. 4GBs).
However, \sys's resource utilization is much lower than Azure (6$\times$ lower for mean CPU usage), indicating that
\revise{its customers may over-provision the hardware resources}.
We identify two possible reasons for that:
(i) \revise{\sys} apps are mostly delay-critical and their usage patterns exhibit high temporal variations, forcing its customers to reserve more resources to ensure a consistently good QoE;
(ii) it remains difficult for edge customers to forecast the fluctuating resource demands at different locations.
Our findings suggest that existing resource allocation challenges are amplified when apps are migrated from clouds~\cite{conf/asplos/DelimitrouK14,meisner2011powernap} to edges.

\textit{(5) \textbf{Resource usage} ($\S$\ref{sec:vm-usage})}
is a critical piece of information that an edge service provider needs to closely keep track of, and
\textit{\textbf{(6) load balancing}} ($\S$\ref{sec:resource-allocate}) facilitates edge services' SLA
by adapting resource allocation to applications' needs.
We find that for \sys, its resource usage is highly unbalanced across servers (up to 14$\times$ from the same \node), across \nodes (up to 731$\times$ in the same province), and across the VMs hosting the same app (up to 3$\times$).
These observations indicate possible imperfections of \sys's VM placement and selection strategies.
We also identify important factors to consider when designing load balancers for \revise{\sys-like} edge platforms, including the fluctuating usage patterns,
the geo-sensitive resource demand, and the decoupled VM placement and end-user request scheduling strategies.
Fortunately, our further experiments of \textbf{\textit{(7) resource prediction}} ($\S$\ref{sec:pred}) show that \sys workloads have stronger seasonality and are easier to predict as compared to Azure.
It offers a good opportunity for more fine-grained, intelligent resource management.


\textit{(8) \textbf{Monetary cost} ($\S$\ref{sec:cost})}
is a critical dimension of commercial edge services but is rarely studied by prior literature.
%
We find that the apps deployed on \sys are mostly bandwidth-hungry, which often constitutes most of their billing cost.
Since data is generated by nearby users, deploying applications over \sys is indeed much cheaper compared to \cloudsys (recall that both are Chinese providers) --
about 45\% of cost reduction on average, and up to 98\% for network cost reduction, making it one of the strongest incentives to move \revise{from cloud to \sys}.
%
However, we also discover that two types of apps may not get financial benefits if deployed on \sys:
(i) apps with high hardware demand but low network demand, as \sys charges slightly higher on hardware resources;
(ii) apps with high temporal network usage variance, as \sys adopts a very coarse-grained billing model.

\noindent \textbf{Contributions}
This work presents a first-of-its-kind measurement study of \revise{a major public edge platform in China}
from both the end users' and the edge operators' perspectives.
%
Our contributions consist of detailed characterizations of the performance, workload, and billing of the edge platform.
Based on our findings, we summarize the key lessons learned in~\S\ref{sec:discuss}.
Given the increasing prevalence of edge computing (in particular fueled by 5G), our work provides crucial insights towards improving future edge services.
Meanwhile, our results also provide an important “baseline” for studying how it evolves in the future.

\revise{
\noindent \textbf{Open source} The edge workloads traces we collected are available at \url{https://github.com/xumengwei/EdgeWorkloadsTraces}.
}

%% file: tab-density.tex
\begin{table}[t]
	\centering \scriptsize
	\begin{tabular}{|l|ll|l|l|ll|l|}
		\hline
		\textbf{\begin{tabular}[c]{@{}l@{}}Plat-\\ form\end{tabular}} & \multicolumn{2}{l|}{\textbf{\begin{tabular}[c]{@{}l@{}}Regions /\\ Coverage\end{tabular}}} & \textbf{\begin{tabular}[c]{@{}l@{}}Density\\ ($10^6 mi^2$)\end{tabular}} & \textbf{\begin{tabular}[c]{@{}l@{}}Plat-\\ form\end{tabular}} & \multicolumn{2}{l|}{\textbf{\begin{tabular}[c]{@{}l@{}}Regions /\\ Coverage\end{tabular}}} & \textbf{\begin{tabular}[c]{@{}l@{}}Density\\ ($10^6 mi^2$)\end{tabular}} \\ \hline
		& 24 & Global & 0.13 &  & 33 & Global & 0.17 \\ \cline{2-4} \cline{6-8} 
		\multirow{-2}{*}{\begin{tabular}[c]{@{}l@{}}AWS\\ EC2\end{tabular}} & 6 & U.S. & 1.58 & \multirow{-2}{*}{\begin{tabular}[c]{@{}l@{}}MS\\ Azure\end{tabular}} & 8 & U.S. & 2.11 \\ \hline
		& 24 & Global & 0.13 &  & 23 & Global & 0.12 \\ \cline{2-4} \cline{6-8} 
		\multirow{-2}{*}{\begin{tabular}[c]{@{}l@{}}Google\\ Cloud\end{tabular}} & 8 & U.S. & 2.10 & \multirow{-2}{*}{\begin{tabular}[c]{@{}l@{}}Alibaba\\ Cloud\end{tabular}} & 12 & China & 3.23 \\ \hline
		
		\begin{tabular}[c]{@{}l@{}}Azure\\ Edge Zones\end{tabular} & 5 & U.S. & 1.32 & {\color[HTML]{000000} \begin{tabular}[c]{@{}l@{}}Huawei\\ Cloud \end{tabular}} & {\color[HTML]{000000} 5} & {\color[HTML]{000000} China} & {\color[HTML]{000000} 1.35} \\ \hline
		
		\begin{tabular}[c]{@{}l@{}}AWS Wav-\\ elength +\\Local Zones\end{tabular} & 14 & U.S. & 3.70 & {\color[HTML]{9A0000} \begin{tabular}[c]{@{}l@{}}\sys (our\\  study)\end{tabular}} & {\color[HTML]{9A0000} \textgreater{}500} & {\color[HTML]{9A0000} China} & {\color[HTML]{9A0000} \textgreater{}135} \\ \hline
	\end{tabular}
\caption{A comparison of \sys's deployment with other popular cloud/edge services. Dated to May. 26, 2021.
}
\vspace{-15pt}
\label{tab:density}
\end{table}


%% file: sys.tex
\section{The \sys Edge Platform}\label{sec:ens}


\textbf{Context and terminology}
The primary differences between \sys and cloud providers, e.g., Alibaba Cloud (\cloudsys) and AWS EC2, are how physical servers are located, organized, and maintained.
While cloud providers also build their large data centers across different geographical locations, edge providers take a step further and treat such geo-distribution as their first-class target.
We call data centers at different locations \textbf{\nodes}.
A \node consists of many \textbf{servers}, and each server hosts many VMs.
The customers of \sys typically subscribe to one or multiple VMs, on which they operate applications or services.
In this study, we assume the VMs that use the same system image and belong to the same user serve the same application (\textbf{edge app}).
Figure~\ref{fig:overview} shows the overall organization of \sys and our measurement methodology as will be discussed in the following subsection.

\input{fig-overview}

\textbf{\sys overview}
While still at its early stage, \sys has now become a leading edge platform in China.
Compared to cloud platforms that typically have less than 10 \nodes in one country, \sys's \node number is about two orders of magnitudes larger and the number is still fast-growing.
Such a difference leads to a significant chain reaction in other aspects of the platforms such as app performance, resource usage, and so on as we will characterize in the following sections.
A \node in cloud computing often hosts thousands or even millions of servers and the number is highly scalable; while a \sys \node typically hosts only tens or hundreds of servers as constrained by the physical infrastructure, e.g., space and electricity.
While \sys supports many types of services (e.g., PaaS and FaaS), the current dominant usage is Infrastructure-as-a-Service (IaaS) VMs.
Thus, this paper mainly targets at IaaS VMs hosted in \sys for workload analysis.
\revise{The physical servers of \sys come from many sources.
The majority of them are built atop Alibaba CDN PoPs.
Some are cooperatively managed by \sys and other third-part IDCs or network operators.
\sys also provides business customers with edge infrastructures that are hosted on the customers' own hardware.
Nevertheless, the current form of \sys is mainly based on micro datacenters and has not generally sunk into cellular core networks as envisioned by MECs~\cite{mec-whitepaper}.
}



\textbf{\sys operation}
Just as cloud, deploying an app on \sys takes two main stages.
(1) \textbf{VM placement by edge provider.}
The customers first submit their resource requirements at different geographical locations to \sys administrators.
For example: ``I need 10 virtual machines in Guangdong province, each with 16 CPU cores and 32GB memory.''
\revise{Generally speaking, \sys only exposes a relatively coarse spatial granularity for customers to subscribe (e.g., province instead of \node).
This is to ensure an elastic resource allocation strategy, as the resources available on each \node are very limited.
}
Once a subscription request arrives, \sys returns one feasible allocation.
While there are often thousands of options, \sys favors the servers that are low in usage in terms of the sales ratio and actual CPU usage (mean and max).
(2)
\textbf{End-user traffic scheduling by edge customers.}
Once \sys allocates the VMs, customers take over the whole control of those VMs.
They are also in charge of scheduling the requests from end users to a given VM.
Similar to traffic routing in content delivery network (CDN), edge customers typically route user requests to their nearby \nodes based on DNS or HTTP 302.

\subsection{Measurement Methodology}\label{sec:data-collection}

We collect two kinds of datasets from \sys:
(i) edge performance ($\S$\ref{sec:data-active}), for which we \textit{actively} build benchmark tools and obtain testing results through crowdsourcing and controlled experiments;
(ii) edge workloads ($\S$\ref{sec:data-passive}), for which we \textit{passively} log the edge VMs' activities and traces.

\subsubsection{Edge Performance Data Collection}\label{sec:data-active}

We actively collected three kinds of data: network latency, network throughput, and application-level performance, for both edges and clouds.
The first two were obtained by crowdsourcing, while the other one is performed in controlled settings.

\textbf{Edge and cloud servers}
(1) For latency,
we set up one VM on each edge \node of \sys and each cloud region of \cloudsys.
Those VMs are used as the ping destinations.
(2) For throughput, we set up 20 \sys VMs at different cities, each with 1Gbps bandwidth capacity.
We didn't use \cloudsys or all \sys regions because the experiments impose too much traffic overhead.
However, the 20 VMs are enough to draw our key conclusions as will be later presented.
(3) For application QoE, we set up 1 nearest edge VM and 3 cloud VMs at different locations that are 670Km/1300Km/2000Km away from where the experiments are performed.
Each VM has 8 vCPUs (2.5GHz), 16GBs memory, and sufficient bandwidth.

\textbf{User equipments (UE)}
We use several commodity off-the-shelf UEs for crowd-sourced network measurements. 
For application QoE testing, we used one laptop (MacBook Pro, 2019 version, 16-inch) and three smartphones: Samsung Note 10+ (Snapdragon 855, 5G-supported), Xiaomi Redmi Note 8 (Qualcomm Snapdragon 665), and Nexus 6 (Qualcomm Snapdragon 805).
We mainly used Qualcomm chipsets because the GamingAnywhere~\cite{gaminganywhere} framework cannot utilize the built-in codec hardware for other chips.

We mainly focus this study on smartphones because (1) smartphone is often regarded as the major type of UE for accessing edge resources, and (2) the wifi speed on smartphones and laptops are similar as we have measured.

\textbf{Testing tools and applications}
We mainly used traceroute (ICMP) and iPerf3 (TCP) to obtain the network latency and throughput performance.
We also built two QoE-testing apps, which are commonly envisioned to be (future) killer apps in the era of  edge computing.
(1) \textit{Cloud gaming}:
we adopted three desktop games (\textit{Battle Tanks}~\cite{game-btanks}, \textit{Pingus}~\cite{game-pingus}, and \textit{Flare}~\cite{game-flare}) to be cloud-powered based on GamingAnywhere~\cite{gaminganywhere}, the state-of-the-art cloud gaming platform.
Edge/cloud servers are to receive player actions from UEs, perform game logic, render the images, and finally encode and send them back to the UE for display.
(2) \textit{Live streaming}:
we built a live streaming app based on real-time messaging protocol (RTMP) with Nginx~\cite{nginx} (server side, Ubuntu), EasyRTMP-Android~\cite{easyrtmp-android} (sender UE, Android device), and MPlayer~\cite{mplayer} (receiver UE, Mac Laptop).
In this application, edge/cloud servers are to pull the videos from the sender UE, (optionally) transcode the videos, and push them to the receiver UE.

\textbf{Testing process}
(1) For latency, we recruited volunteers in China using Android devices. We installed our speed-testing app on their devices, and asked them to run the tests.
During testing, the app will obtain the round-trip time (RTT) to each edge/cloud VM we set up and the intermediate hops if visible.
Each IP testing is repeated by 30 times.
Once finished, the testing results will be encrypted and uploaded to our server, along with the network condition (WiFi/LTE/5G), testing time, and the city name.
In total, we received 385 testing results ($>$2M pings) from Jun. 1st to Aug. 1st in 2020\footnote{The volunteers recruited are not affiliated with \sys.
All participants are paid for their efforts and the traffic data consumed in the experiments.}.
The results come from 158 users, covering 20 provinces, and 41 cities in China.
For network type, 59\%/34\%/7\% of the testings are performed under WiFi/LTE/5G.
\revise{During each test, we ask the participants to keep their smartphones in a stationary context, e.g., no WiFi/4G switching or 4G handoff.
This is ensured by our testing script that monitors the network condition and physical motions of the devices.
}
(2) For throughput, we selected 25 volunteers at different cities, a subset from the above, to run our testing script.
The script used iPerf3 to get both downlink/uplink throughput to each of the 20 edge VMs we selected, where iPerf3 runs for 15 seconds per connection.
(3) The application QoE experiment was performed by the authors.
Each testing was repeated across 4 different locations in the same city: campus indoor/outdoor and office building indoor/outdoor.

\subsubsection{Edge Workloads Data Collection}\label{sec:data-passive}
This dataset contains information about every VM
running on \sys from June 1st to Sep 1st, 2020.
More specifically:
(1) a VM table, with each VM's placement information (which server and \node it's hosted at), customer information (whom it belongs to), and system information (the image id, os type, kernel number, etc);
(2) the resource size (capacity) in terms of maximum CPU cores, memory, and disk for each VM and server;
(3) the CPU usage reported every 1 minute for each VM;
(4) the bandwidth usage reported every 5 minutes for each VM, including both private (intra-\node) and public traffic.

\subsection{Selecting cloud workloads for comparison}\label{app:traces}

\input{tab-traces}

\revise{
The goal of this work is to compare \sys with cloud platforms to reveal their disparity, and therefore showcase the key benefits brought by \sys.
Regarding the workloads comparison ($\S$\ref{sec:workloads}), we investigate the cloud workloads datasets that are publicly available and summarize them in Table~\ref{tab:traces}.
Next, we describe these datasets in detail and explain why we choose to use or not use each of them for comparison.
}

\begin{myitemize}
	\item \textbf{Azure dataset}~\cite{cortez2017resource} is the most representative counterpart of \sys on public cloud platforms and thus comprehensively compared in this work (we used the 2019 version).
	\item \textbf{AliCloud dataset}~\cite{alibaba-cloud-data} is not compared because: (1) It only contains the usage of containers instead of VMs, \revise{while the major form of \sys is VM}; (2) Its time range is too short for certain analysis (8 days), e.g., resource usage profiling and prediction.
	\item \textbf{Google dataset}~\cite{borg} is not compared because: (1) Its resources are not available to public but only to Google's internal developers, making it not representative of public cloud platforms.
	(2) The dataset access is through Google's BigQuery interface, which doesn't support complicated usage such as ML-based prediction.
	\item \textbf{GWA-T-12 dataset}~\cite{shen2015statistical} is not compared because it's too small-scale and out-of-date.
\end{myitemize}

\subsection{Ethics}\label{app:ethics}
\revise{
When conducting this study, we take careful steps to protect user privacy and preserve the ethics of research.
(i) For collecting the edge performance dataset,
the data collection was approved by the Research Ethical Committee of the institutes that the authors are currently affiliated with;
the collection was also approved by the participants ahead of experiments through informed consent; 
we collected no sensitive data from the participants except their residential city, which was input by the participants themselves.
(ii) For collecting the edge workloads dataset of \sys,
the data collection was approved by its customers through the service agreement;
no customer identifiable information was collected during the study.
When exported, the customer ID of the dataset is anonymized.
}

%% file: fig-overview.tex
\begin{figure}
	\centering
		\includegraphics[width=0.48\textwidth]{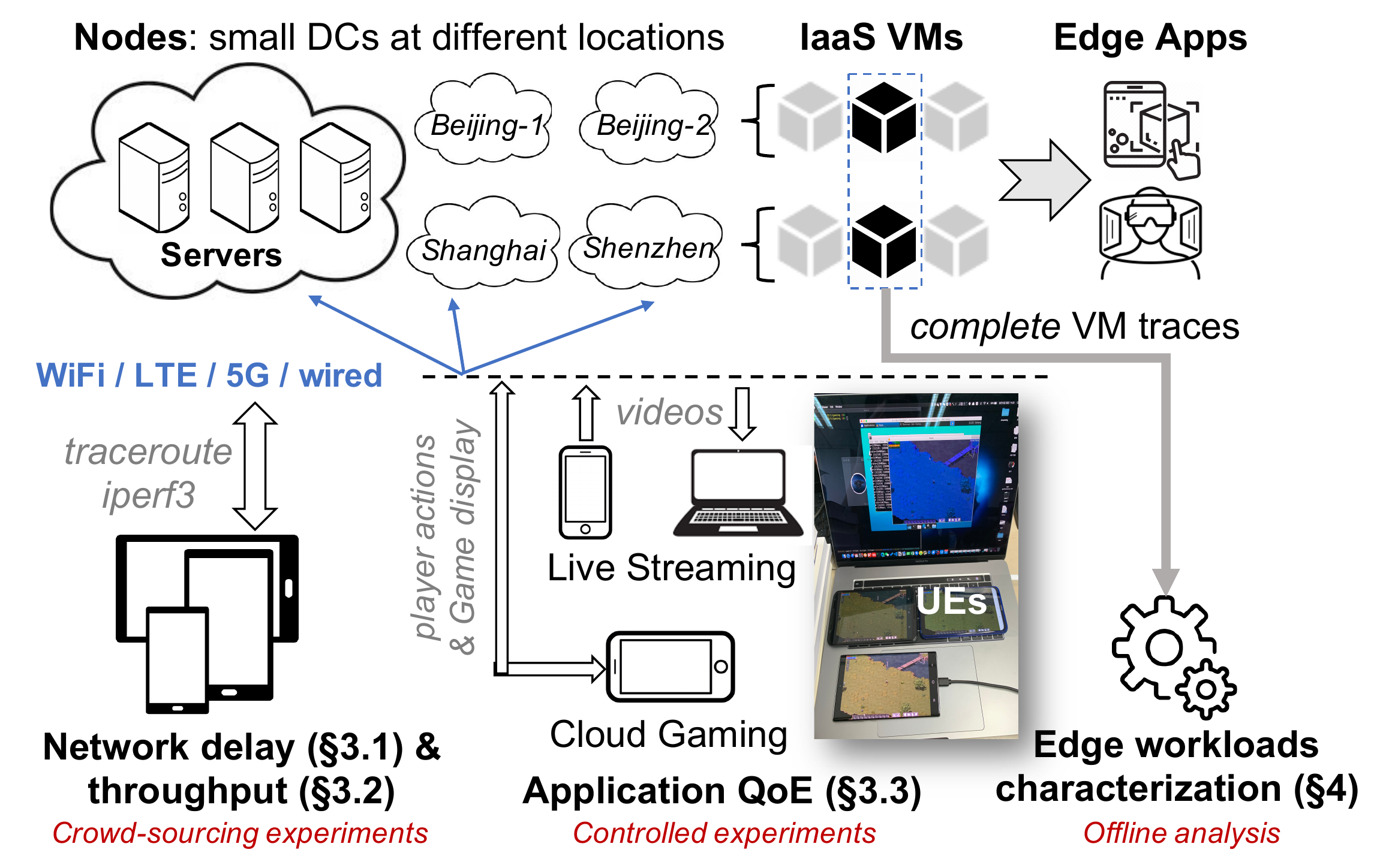}
		\caption{\textbf{The overall organization of \sys platform (top) and our measurement methodologies (bottom).}}
		\label{fig:overview}
\vspace{-15pt}
\end{figure}

%% file: tab-traces.tex
\begin{table*}[]
\footnotesize
\begin{tabular}{|l|l|l|l|l|l|}
\hline
\textbf{} & \textbf{Platform} & \textbf{Duration} & \textbf{Scale} & \textbf{Customers} & \textbf{Why it's not compared?} \\ \hline
\textbf{Azure Dataset}~\cite{cortez2017resource} & Azure Cloud & \begin{tabular}[c]{@{}l@{}}1 month in 2017\\ 1 month in 2019\end{tabular} & \begin{tabular}[c]{@{}l@{}}2.0M VMs\\ 2.7M VMs\end{tabular} & public & The 2019 version is used. \\ \hline
\textbf{AliCloud Dataset}~\cite{alibaba-cloud-data} & AliCloud ECS & \begin{tabular}[c]{@{}l@{}}12 hours in 2017\\ 8 days in 2018\end{tabular} & \begin{tabular}[c]{@{}l@{}}1.3k servers\\ 4.0k servers\end{tabular} & public & Only containers' usage are included. \\ \hline
\textbf{Google Dataset}~\cite{borg} & Google Borg & \begin{tabular}[c]{@{}l@{}}1 month in 2011\\ 1 month in 2019\end{tabular} & \begin{tabular}[c]{@{}l@{}}12.6k servers\\ 96.4k servers\end{tabular} & Google developers & Only support BigQuery. Not public platform. \\ \hline
\textbf{GWA-T-12}~\cite{shen2015statistical} & Bitbrains & 3 months in 2013 & 1.75k VMs & Enterprises & Old. Not publicaly available. Small scale. \\ \hline
\textbf{Our Dataset} & NEP & 3 months in 2020 & Complete set & public & / \\ \hline
\end{tabular}
\caption{A comparison of cloud/edge workloads traces that are publicly available. We also explain why we choose Azure as the cloud-side counterpart for head-to-head comparison in this work.}
\label{tab:traces}
\vspace{-10pt}
\end{table*}

%% file: performance.tex
\section{Demystifying Edge Performance}\label{sec:performance}

\input{measurement-network-latency}

\input{measurement-throughput}

\input{measurement-app-perf}

%% file: measurement-network-latency.tex
\subsection{End-to-end Network Latency}\label{sec:network-delay}
Based on the collected data ($\S$\ref{sec:data-active}), we first \revise{calculate the median network delay among each user and \sys \node}, and then aggregate the results across users. This is to eliminate the impacts from heavy users who have run our testing multiple times.

For simplicity, we define the ``nearest edge/cloud'' as the edge/cloud \node that has the smallest \revise{median} RTT to an end user.
Besides the nearest edge/cloud that represents the optimal network performance available in the current deployment of \sys/\cloudsys, we include two other baselines:
(i) the 3rd-nearest edge, for which we will show that there are multiple edges that are close to each user;
(ii) all clouds, which is averaged across all the \nodes of \cloudsys. 
\revise{This baseline reflects the performance of deploying on a centralized server for users of a nation (China in our case), a common tradeoff among economic and performance perspectives.}

\input{fig-rtt}

\textbf{Overall RTT}
Figure~\ref{fig:rtt}(a) illustrates the \revise{median} RTTs across users under different network types.
Under WiFi, the median RTT for the nearest edge \node is \revise{10.5ms}, which is \revise{1.89$\times$ (19.8ms)} faster than the nearest cloud \node, and \revise{3.4$\times$ (35.7ms)} faster than all clouds on average.
The 3rd nearest edge \node also provides smaller network latency \revise{(15.5ms)} than the nearest cloud.
Under LTE, the overall improvement decreases: the median latency for the nearest edge is \revise{34.2ms, which is only 1.42$\times$/1.93$\times$} faster than the nearest/all cloud \nodes.
The decreased latency reduction comes from that the first 2 hops of LTE network incur much higher latency than WiFi (47.8ms vs. 9.0ms on average).
We will give more details of hop-level latency breakdown later in this section.

For 5G\footnote{In China, 5G network operates at 3.5 GHz frequency.
Note that comparing 5G to LTE is not the focus of this study, for which we refer readers to~\cite{5g-measurement,5g-mmwave}.}, the median RTT of the nearest edge is only 10.4ms.
Its significant improvement over LTE mainly attributes to the flatten architecture of 5G and the improved fiber fronthaul/backhaul~\cite{3gpp-2019,5g-measurement}.
The improvement over all clouds is also tremendous \revise{(2.64$\times$)}.
However, the improvement is much smaller \revise{(1.35$\times$)} compared to the nearest cloud.
We dig into our trace and find out that almost all our 5G testing results are from Beijing due to very limited 5G coverage in other regions in China.
Since \cloudsys also deploys a \node in Beijing, the difference in accessing \sys and \cloudsys is trivial.
We expect the network improvement brought by \sys to be more significant when 5G infrastructures become more widely deployed and accessible to more end users.

\revise{We also analyze the average RTT and physical distance to the nearest edge/cloud \node across users based on their locations, i.e., whether they are co-located with an edge/cloud \node in the same city.
The results are summarized in Table \ref{tab:rtt-impr-per-user}. 
In our experiments, most users (69\%) are not co-located with any edge/cloud \node.
In such a circumstance, the average RTT is reduced from 34.97ms (to the nearest cloud) to 22.37ms (to the nearest edge) with \sys, and the geographical distance is reduced from 351km to 130km.
The reduction is much more significant when the users are co-located with a \sys \node but not a cloud 
\node (18\%), i.e., 47.06ms to 18.45ms.
For cases where users are co-located with both edge/cloud \nodes, we find \sys edge can still improve the RTT.
The reason is that \sys deploys multiple \nodes in a few cities, e.g., Beijing, so that the users in those cities can access nearer resources.
In summary, while \sys delivers lower network delay to end users through resources in proximity, the benefits vary across different locations of endpoints.
}

\textbf{Network jitter}
Many network-sensitive tasks like live streaming are required to deliver consistent, predictable user experience.
To quantify the network jitter, we measure the RTT coefficient of variation (CV for short, measured as stddev/mean) during our repetitive tests (30 times) for each experiment.
As illustrated in Figure~\ref{fig:rtt}(b), edge platform has significantly lower RTT CV (i.e., higher stability) compared to cloud platform.
Under WiFi/LTE/5G, the median RTT CV is only 1.1\%/2.3\%/0.7\% for the nearest edge and 1.5\%/3.2\%/1.7\% for the 3rd nearest edge.
Taking the nearest edge as baseline, the nearest cloud \node has 5.8$\times$/3.9$\times$/5.7$\times$ higher median RTT CV, and the average numbers across all \nodes can be up to 30$\times$.
Such a low network jitter is critical to provide service-level agreement (SLA) to edge customers.

\input{fig-first-hop}

\textbf{Per-hop latency breakdown}
Table~\ref{tab:first-hop} illustrates the hop-level breakdown of the end-to-end RTT.
We highlight the latency of the first 3 hops and combine the rest.
For WiFi, the first wireless hop contributes to 44.2\%/30.2\% of the end-to-end latency to the nearest edge/cloud.
For LTE, the second hop contributes the most latency, e.g., 70.1\% to the nearest edge.
\revise{This is because the 2nd hop contains the network delay \textit{accumulated} from multiple physical hops in the GTP-U tunnel, where data packets are encapsulated in GTP Protocol Data Units and the hop count is not changed during the transmission~\cite{3gpp}.
Therefore, the ``hop'' latency is longer.
Such an observation is consistent with a recent measurement study~\cite{5g-measurement}.}
For 5G, our collected trace doesn't contain the latency of the first 2 hops, possibly because the ICMP service is disabled by the operator.
Instead, we report the latency of the first 3 hops in total, and find they dominate the end-to-end latency, i.e., 98\% for the nearest edge.
Note that, compared to cloud platforms, the current deployment of \sys mainly reduces the inter-city transmission delay (i.e., the backbone network).
The traffic still needs to travel through the core network within a city to reach the edges.

\textbf{Hop number}
Figure~\ref{fig:hop-num} illustrates the number of hops between end devices and edge/cloud servers, averaged across all network types.
It shows that the hop number to the nearest edge (5--12) is much fewer than the clouds (10--16).
The reduced hop number leads to lower network latency and jitter.
To further reduce the hop distance, \sys needs to increase the \node density and sink the resources into the core network by collaborating with operators as aforementioned.

\textbf{Inter-\node RTT}
We also measure the network latency between \sys's \nodes.
We obtain the RTT between every \node pair every 5 minutes in a day of June 2020, and average the results.
Figure~\ref{fig:inter-node-rtt} illustrates the geographical distances (x-aixs) and network latency (y-axis) between edge \nodes.
Overall, the RTTs increase with the inter-\node distances, and reach 100ms when two \nodes are 3000km away.
More importantly, it shows there are many nearby edge \nodes that have very low RTT, thanks to the deployment density of \sys.
For each \node, there are \revise{1/3/11} nearby \nodes that are within 5ms/10ms/20ms RTTs on average.
It promises fine-grained resource and user request scheduling between edge \nodes.

\implications{
\sys delivers noticeably lower and more stable network delay for end users compared to \cloudsys.
Despite that, \sys hasn't fully reached the envisioned prospects of edge computing (even with current 5G), e.g., sub-10ms delay and 1--2 hops distance to access edge resources~\cite{mec-whitepaper}.
The last-mile hops (1st for WiFi and 2nd for LTE) become the bottleneck of network delay.
To step forward, \sys needs to deploy denser \nodes and collaborate with operators to sink the edge resources into ISP's core networks or even cellular base stations, i.e., Mobile Edge Computing~\cite{mec-whitepaper,aws-wavelength}.
}

\input{tab-rtt-impr-per-user}

%% file: fig-rtt.tex

\begin{figure}[t]
	\centering					
	\begin{minipage}[b]{0.45\textwidth}
		\includegraphics[width=1\textwidth]{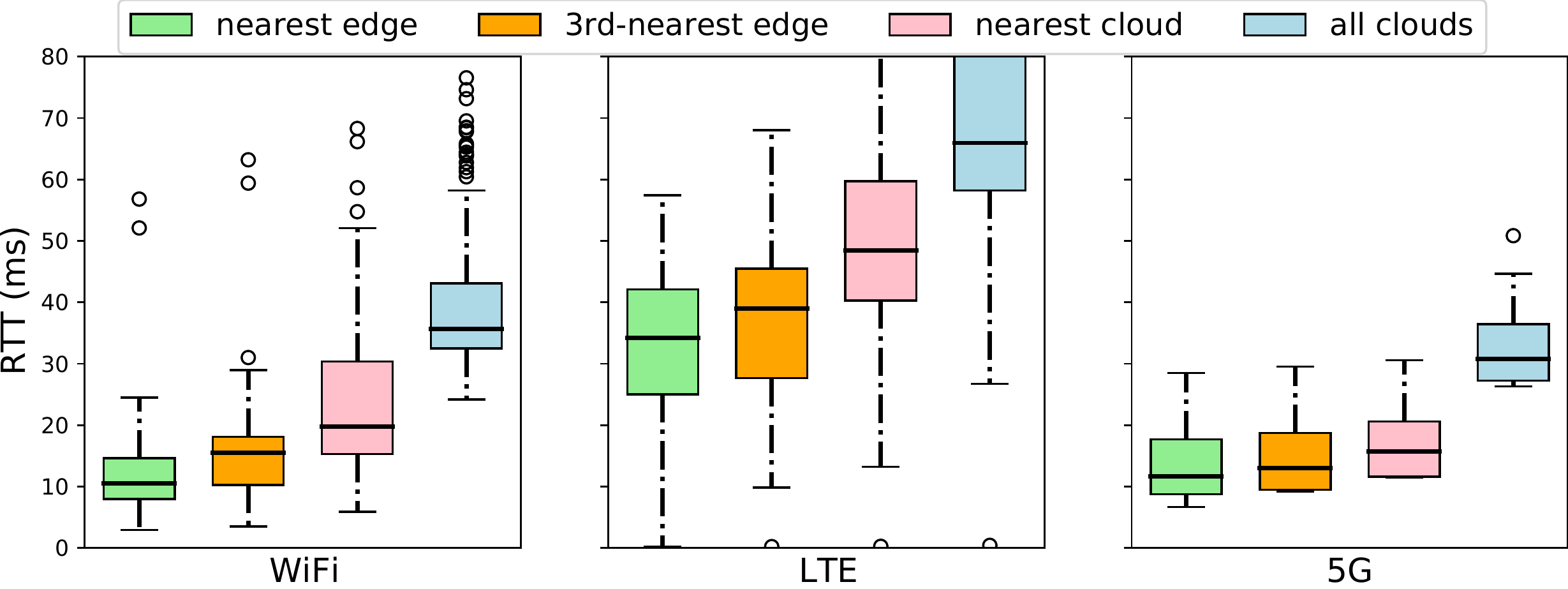}
		\subcaption{\revise{Median} RTT across users.}
	\end{minipage}	
	
	\begin{minipage}[b]{0.45\textwidth}
		\includegraphics[width=1\textwidth]{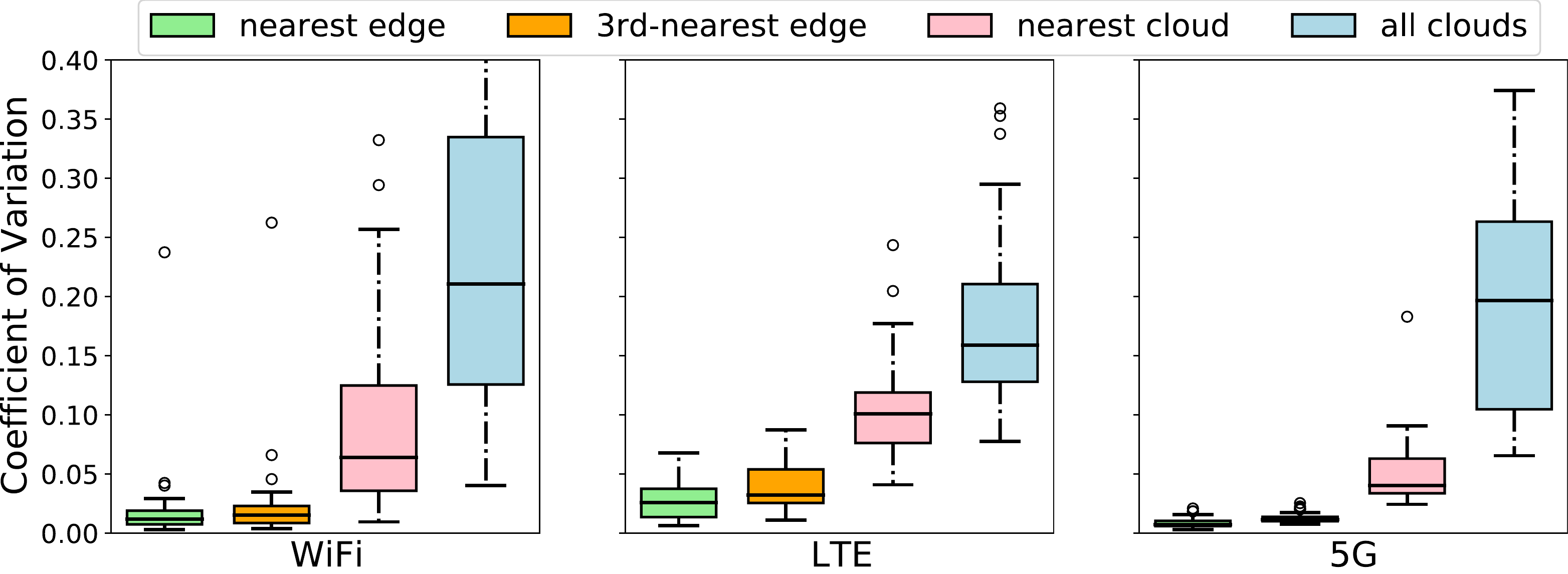}
		\subcaption{RTT coefficient of variation (CV) across users.}
	\end{minipage}
	\caption{\textbf{The network delay (median RTT) and jitter (RTT CV)} from end users to edge/cloud \nodes.}
	\label{fig:rtt}
	\vspace{-10pt}
\end{figure}

%% file: fig-first-hop.tex

\begin{table}[t]
\footnotesize
\centering
\begin{tabular}{|l|l|l|l|l|}
	\hline
	\multirow{2}{*}{} & \multicolumn{2}{c|}{\textbf{Nearest edge \node}} & \multicolumn{2}{c|}{\textbf{Nearest cloud \node}} \\ \cline{2-5} 
	& 1st-2nd-3rd hop & Rest & 1st-2nd-3rd hop & Rest \\ \hline
	\textbf{WiFi} & 44.2\%-10.3\%-15.1\% & 30.2\% & 30.1\%-5.0\%-11.5\% & 52.5\% \\ \hline
	\textbf{LTE} & 10.2\%-70.1\%-9.4\% & 10.3\% & 10.1\%-51.6\%-13.1\% & 25.2\% \\ \hline
	\textbf{5G} & 97.9\% in total & 2.1\% & 82.2\% in total & 17.8\% \\ \hline
\end{tabular}
\caption{\textbf{Hop-level breakdown of network delay}}
\label{tab:first-hop}
\vspace{-15pt}
\end{table}


\begin{figure}[t]
	\centering					
	\begin{minipage}[b]{0.23\textwidth}
		\includegraphics[width=1\textwidth]{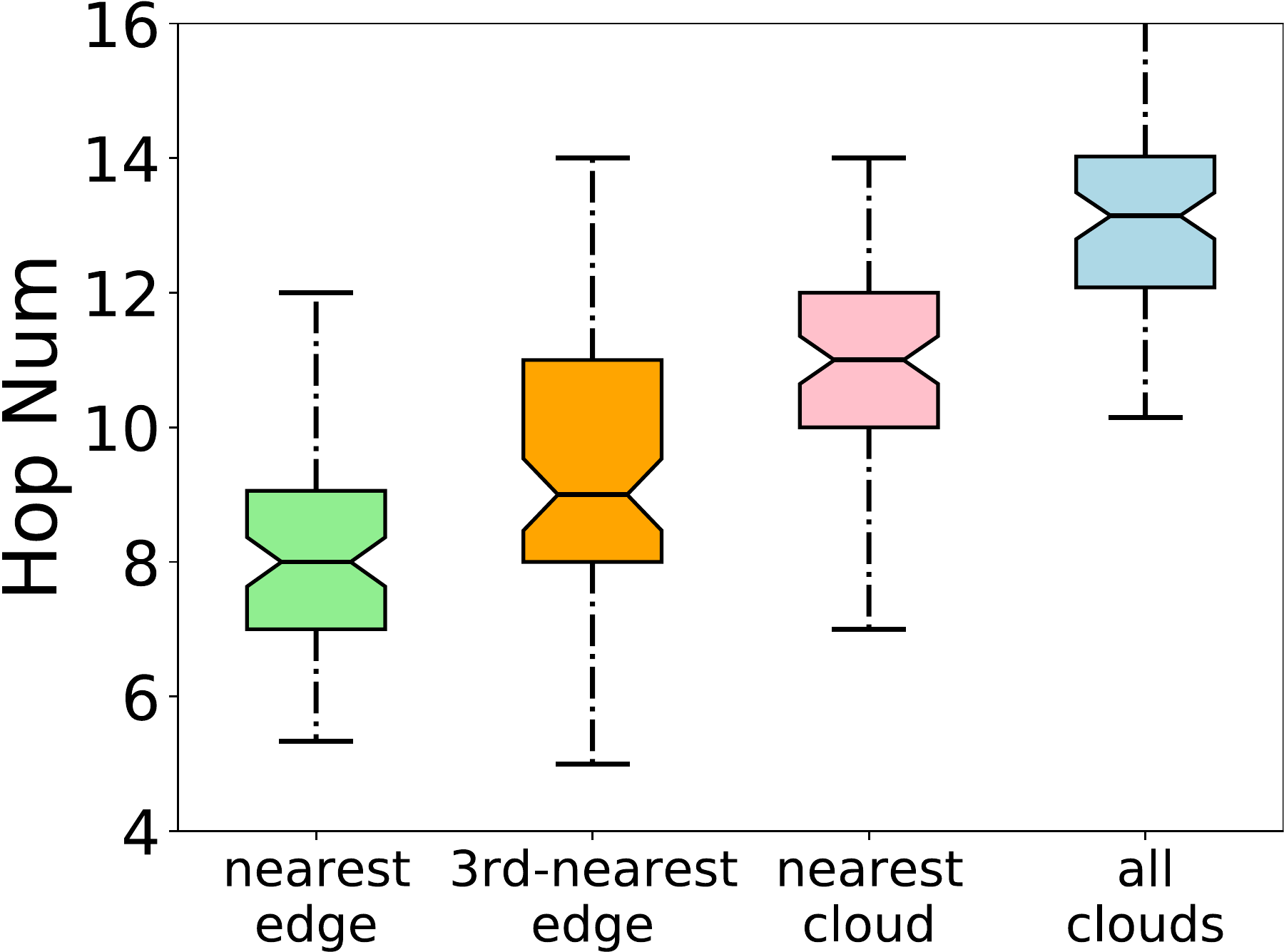}
		\caption{\textbf{Hop numbers}}
		\label{fig:hop-num}
	\end{minipage}\vspace{-10pt}
	~
	\begin{minipage}[b]{0.23\textwidth}
		\includegraphics[width=1\textwidth]{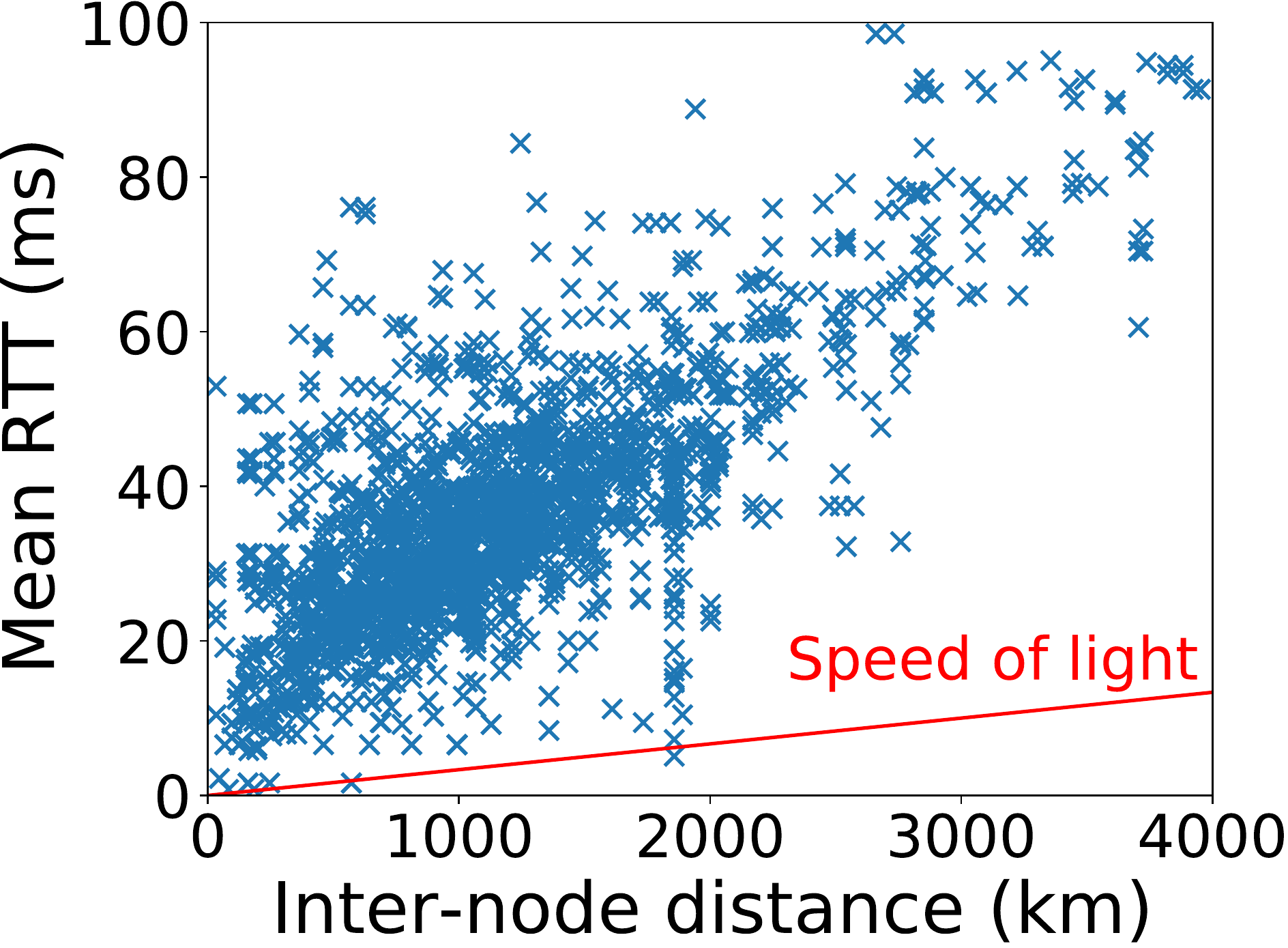}
		\caption{\textbf{Inter-\node RTTs}}
		\label{fig:inter-node-rtt}
	\end{minipage}
\vspace{-5pt}
\end{figure}

%% file: tab-rtt-impr-per-user.tex
\begin{table}[]
    \footnotesize
    \begin{tabular}{|c|c|c|c|c|}
    \hline
    \multicolumn{1}{|c|}{\textbf{\begin{tabular}[c]{@{}c@{}}U/E/C\\ Locations\end{tabular}}} & \multicolumn{1}{c|}{\textbf{\begin{tabular}[c]{@{}c@{}}RTT (ms)\\ Nearest-E\end{tabular}}} & \multicolumn{1}{c|}{\textbf{\begin{tabular}[c]{@{}c@{}}RTT (ms)\\ Nearest-C\end{tabular}}} & \multicolumn{1}{c|}{\textbf{\begin{tabular}[c]{@{}c@{}}Dist (KM)\\ Nearest-E\end{tabular}}} & \multicolumn{1}{c|}{\textbf{\begin{tabular}[c]{@{}c@{}}Dist (KM)\\Nearest-C\end{tabular}}} \\ \hline
    U/E \& U/C co-located (13\%)                                                                                       & 10.96                                                                                          & 15.64                                                                                          & 0                                                                                               & 0                                                                                                \\ \hline
    U/E co-located
    (18\%)                                                                                         & 18.45                                                                                         & 47.06                                                                                          & 0                                                                                               &
     973                                                                                    \\ \hline 
    None co-located
    (69\%)                                                                                            & 22.37                                                                                         & 34.97                                                                                           &
    130                                                                                    & 
    351                                                                                     \\ \hline
    \end{tabular}
    \caption{\revise{Average RTT and physical distance to the nearest edge/cloud server across different locations of users.
    ``U/E/C'': user, edge \node, cloud \node.
    ``co-located'' means the user is in a city where at least one edge/cloud \node is deployed.
    When calculating the distance, we look at the geographic distance at city level. Numbers averaged across WiFi/4G/5G.}}
    \label{tab:rtt-impr-per-user}
    \end{table}

%% file: measurement-throughput.tex
\subsection{End-to-end Network Throughput}\label{sec:throughput}

Between a cloud/edge VM and a client, the network throughput is bounded by the poorest link among them, e.g., the first hop for wireless access as commonly believed.
Based on the data collected through crowdsourcing ($\S$\ref{sec:data-active}), this section dives into the question: how does the geographic distance (or hop number) affect the network throughput between end users and data centers.
By answering this question, we can learn whether or how edge platforms like \sys can improve the network throughput compared to remote clouds.

\input{fig-bw}

Figure~\ref{fig:bw-test} illustrates the overall results of our throughput testing.
The key observation is that, when throughput is low e.g., $\le$100Mbps for LTE and WiFi, the correlation between the distance and throughput is negligible, as indicated by Pearson correlation coefficient lower than 0.2~\cite{pearson}.
Noting that the 5G uplink bandwidth (mean: 52Mbps) is strictly capped by asymmetric time slot ratio in the ISP's configuration following Rel-15 TS 38.306~\cite{3gpp}, thus its correlation with distance is also negligible.
Only when the throughput reaches high, e.g., for 5G downlink (mean: 497Mbps) and wired access (mean: 480Mbps), the correlation becomes significant (corr$>$0.7).
In such cases, the throughput degrades observably as physical distance increases.
The reason is that, with LTE/WiFi access, the network throughput is usually bounded by the bandwidth capacity at wireless hop, therefore has little correlation with the distance.
When the capacity is high, e.g., for 5G downlink, the bottleneck resides at the Internet link which directly correlates with the distance (or RTT~\cite{mathis1997macroscopic}).
It is also confirmed by our observations of the TCP congestion window size and the packet loss rate during experiments.

Note that, to have perceivable benefits from the geographically closer edge resources, two more factors need to be satisfied besides the high bandwidth capacity:
(1) Applications that can generate high-volume traffic at more than 200Mbps.
We find that few today's applications can do that: for example, streaming video at 4K resolution and 60FPS consumes only less than 100Mbps~\cite{jigsaw}.
(2) Equally-high or even higher bandwidth needs to be allocated to the edge VMs so that the DC gateway doesn't become the bottleneck.
Such high bandwidth usage, however, can be prohibitively expensive to developers.

\implications{
Bringing resources closer to users improves network throughput \revise{on \sys} only with high bandwidth capacity at the last mile, e.g., for 5G downlink and wired access.
Such an advantage over cloud computing, however, is weakened by the absence of ultra-bandwidth-hungry applications and the cost considerations from developers' perspectives.
Given that, we conclude that improving network throughput is not a primary incentive of current edge applications \revise{on \sys}.
In the future, however, we believe that the throughput improvement will benefit more emerging, bandwidth-hungry edge applications.
}

%% file: fig-bw.tex

\begin{figure}[t]
	\centering					
	\begin{minipage}[b]{0.24\textwidth}
		\includegraphics[width=1\textwidth]{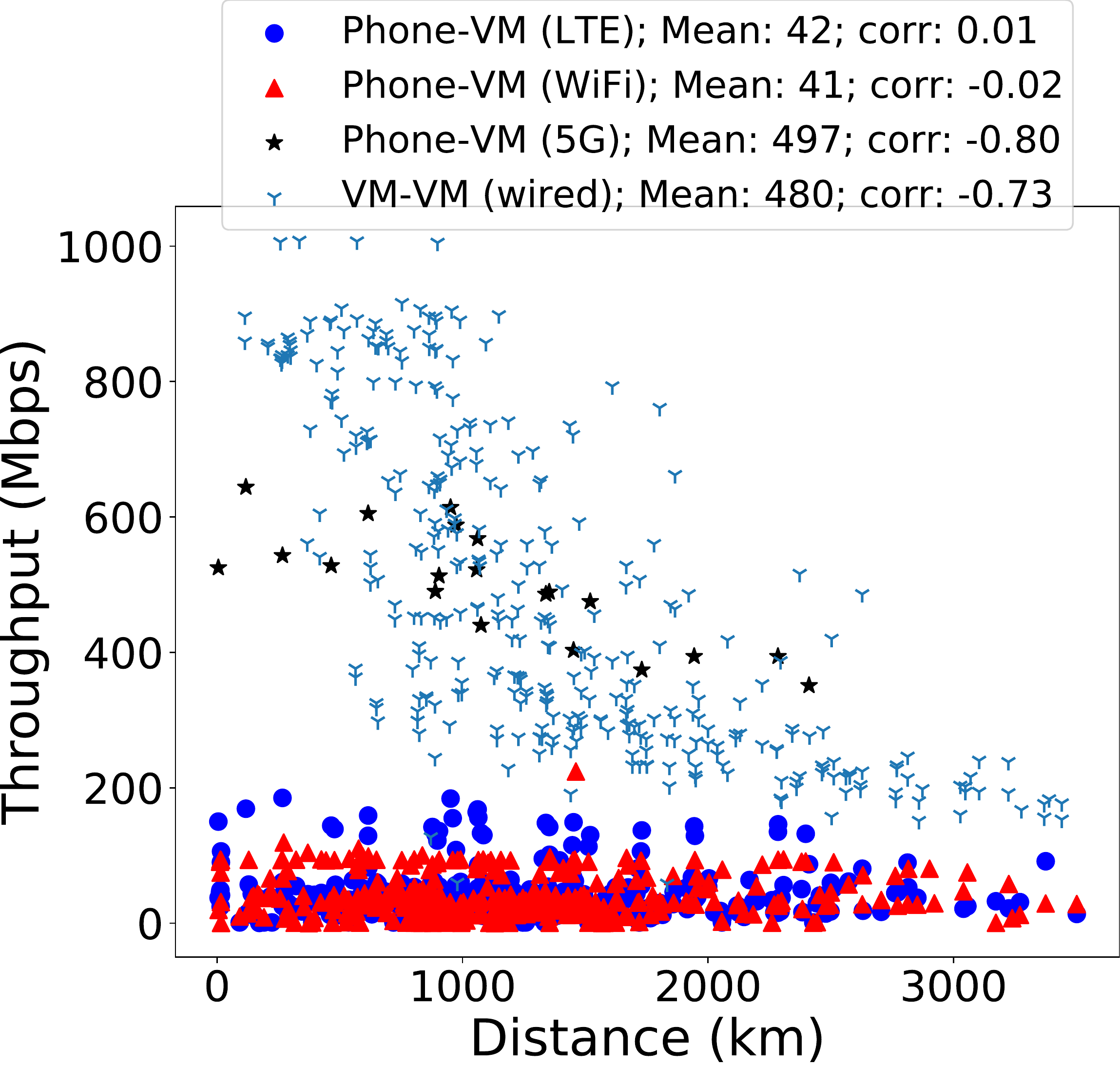}
		\subcaption{Downlink throughput}
		\label{fig:bw-downlink}
	\end{minipage}	
	~
	\begin{minipage}[b]{0.24\textwidth}
		\includegraphics[width=1\textwidth]{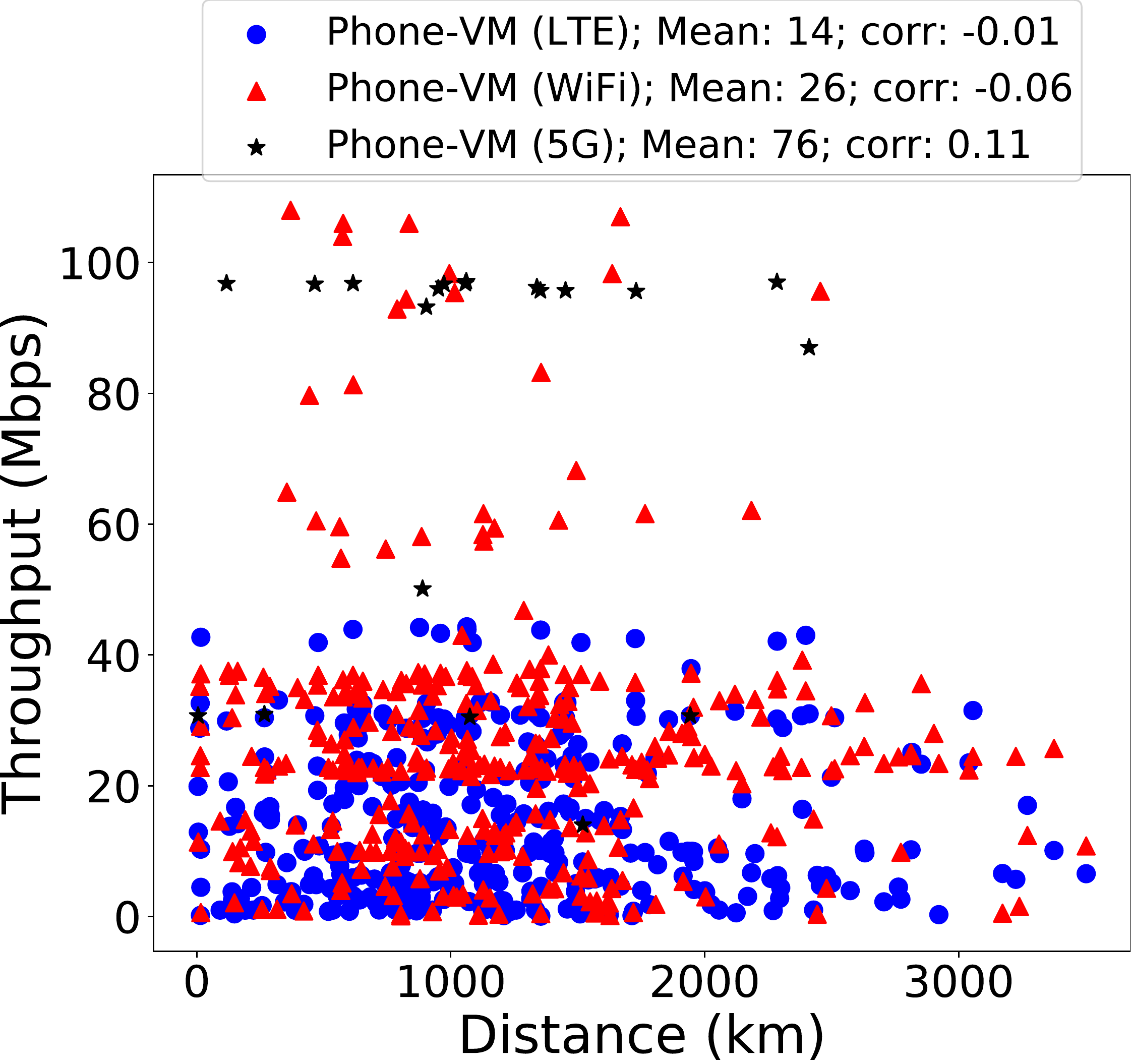}
		\subcaption{Uplink throughput}
		\label{fig:bw-uplink}
	\end{minipage}
	
	\caption{The TCP-based network throughput against geographical distance. Each point represents a  15-sec iPerf-tested result. ``corr'' is the Pearson correlation coefficient (-1 to 1) between distance and throughput.
	}
	\vspace{-10pt}
	\label{fig:bw-test}
\end{figure}

%% file: measurement-app-perf.tex
\subsection{Application Performance (QoE)}\label{sec:app-performance}

\input{tab-app-exp}


This section presents the experiment results of application-level QoE.
Recall that ($\S$\ref{sec:data-active}) we use one nearest edge VM and 3 cloud VMs with different distances from the area where the experiments are carried out.
For reference, Table~\ref{tab:app-perf} shows the average RTTs to those VMs in this experiment.
\revise{
For simplicity, we term the tests as ``Cloud-1/2/3'' from the nearest to the farthest. Note that the locations and resource characteristics of the servers are described in $\S$\ref{sec:data-active}.
}

\subsubsection{Cloud Gaming}\label{sec:cloud-gaming}
By hosting game execution and rendering on backend servers, cloud gaming promises mobile devices the ability to play games at a lower cost~\cite{outatime}.
Cloud gaming systems have stringent response delay requirements, as gamers may demand less than 100ms response delay~\cite{claypool2006latency}.
With cloud servers as the backend, it is difficult for players to attain real-time interactivity in the face of wide-area network latency.
With much lower latency, edge computing is expected to significantly improve the game experience~\cite{5g-cloudgaming}.
We now validate our cloud gaming systems built with edge/cloud backend.

\textbf{Metrics} We follow prior work~\cite{gaminganywhere} to measure the end-to-end performance as the interval between a player issuing a command and the in-game action appearing on the client.
We refer to this metric as \textit{response delay}.
To obtain this delay, we periodically send touch events to the game menu and wait for the menu window to pop up.
In the meantime, we use another device to record the device screen at a high frequency (FPS).
At offline, we use FFmpeg~\cite{ffmpeg} to decode the videos and manually obtain the interval between a touch event being invoked and the menu popping up.
The response delay is thus obtained as the interval between the two timestamps.
The touch event is set to be visible through Android built-in tools.
The online experiment is performed automatically through Android Monkey tool~\cite{android-monkey}.
For each testing, we obtain 50 data points.

\input{fig-game}

\textbf{Overall results}
Figure~\ref{fig:cloud-gaming} shows the cloud gaming performance with different network conditions (a), client devices (b), and game types (c).
Overall, with a good network condition (WiFi in our case) and nearby VMs (Edge and Cloud-1), cloud gaming can achieve less than 100ms response delay.
The distance matters: remote cloud VMs lengthen the delay by up to 60ms.
Among the devices, Samsung Note 10+ performs slightly better than others for its high-end chipset, but the improvement is not significant because the hardware-accelerated video decoding is fast enough for all the devices tested, i.e., less than 10ms for the default 800$\times$600 resolution used in GamingAnywhere, and the screen fresh rate is the same (60Hz).
Note that decoding is needed as the gaming server encodes the game scenes to video frames before sending to UE.
Among different games, \textit{Pingus} experiences slightly higher delay and jitter for its more complex game logic.

\textbf{Breakdown}
An intuitive question is how we can further improve the 100ms response delay, e.g., within 30ms or 2-3 frames.
Our breakdown results show that, the network delay of the nearest edge VM is no longer the bottleneck: the propagation delay is only 11.4ms shown Table~\ref{tab:app-perf} and the transmission delay (to send a frame) is less than 10ms according to our measurement.
Instead, we find the major portion is on the server-side (game logic execution and rendering), which contributes to around 70ms delay.
We further look into this delay and make two interesting observations:
(1) While each server VM is equipped with 8 CPU cores, most cores except one run at very low utilization (less than 10\%). In other words, increasing CPU cores won't help as it is difficult to parallelize the game logic.
(2) Our offline experiment that hosts the server on a Macbook laptop suggests that enabling the GPU rendering can help reduce the delay by around 10ms--20ms.

\implications{
Placing gaming backend on nearby \revise{\sys edges} can help achieve less than 100ms response delay.
To further enhance the experience, we need to improve the server-side gaming execution.
More specifically, adapting the gaming to multi-core systems with high parallelism and applying hardware acceleration (e.g., GPU) are promising approaches.
}

\subsubsection{Live Streaming}\label{sec:live-streaming}
A report~\cite{streaming-report} in 2018 shows that 42\% of the population in the U.S. have now live-streamed online content.
Live streaming also demands low end-to-end delay, especially for bidirectional streaming scenarios that involve human-to-human interaction, e.g., online meeting.
Our workload analysis in $\S$\ref{sec:vm-type} shows that live streaming is one major application type served by \sys.
In this experiment, we assume the sender and receiver are located in the same city, which is common for many streaming scenarios such as online education.
Unless otherwise specified,
we stream 1080p video without transcoding, and the encoded streaming bitrate is around 5Mbps.

\textbf{Metrics}
Following prior work~\cite{5g-measurement}, we define \textit{streaming delay} as the amount of time between a real-world event and the display of that event on the receiver’s screen.
To obtain this delay, we use the sender UE to capture a millisecond-level clock (displayed by a third device), and use a fourth device to capture the running clock and the receiver UE's screen simultaneously.
We then inspect the difference between the two clocks as the streaming delay.
Each testing runs for 20 seconds from which we obtain 50 data points.

\input{fig-streaming}

\textbf{Overall results}
Figure~\ref{fig:live-streaming} shows the live streaming performance under different conditions.
Here, ``WiFi-trans'' indicates transcoding videos from 720p to 1080p on server, while others simply stream videos without transcoding.
We draw the following important observations from this figure.
(1) Edge servers have limited benefit in reducing the streaming delay, e.g., upmost 24\% compared to the farthest cloud under 5G, mostly because the network doesn't constitute the bottleneck as we will show next.
(2) Streaming images with a lower resolution can reduce the delay around 67ms (26\%) from 1080p to 720p.
Note that this reduction not only comes from the reduced network transmission time, but also the rendering on the receiver UE.
(3) Transcoding incurs a high overhead: around 400ms (2$\times$) from 1080p to 720p under our WiFi condition.
This overhead includes both the transcoding time and server waiting time for a video segment to arrive.

All above experiments are carried out without using a \textit{jitter buffer} on the receiver side.
A jitter buffer is commonly used in video streaming to compensate transmission impairments caused by the time-variant packet delays~\cite{huang2014buffer,liang2008effect}.
Our additional experiments show that, with a small jitter buffer (e.g., 2MBs), the streaming delay reaches as high as 2 seconds and the difference between edge/clouds becomes trivial.

\textbf{Breakdown}
Even without using jitter buffer and transcoding, the streaming delay remains about 400ms.
Following the approach in~\cite{jigsaw}, we break down this delay to image capture, local frame processing (frame patch splice, codec, and video rendering), and RTMP network transmission delay.
We draw the following findings.
(1) Network delay takes around 50ms, which doesn't constitute the major bottleneck.
Note that edge reduces only the propagation delay but not transmission delay.
The observation is confirmed by another micro experiment, where we repeat the above experiments but deploying the server on a laptop wired to sender/receiver UEs (LAN environment). In such settings, the streaming delay is only reduced by 40ms.
(2) By calculating the timestamp difference between the running clock and the sender UE's screen, we estimate the image capture plus rendering to be around 140ms.
This delay is sophisticated, including the digital processing by camera image signal processor and the time spent on the system-software stack (Android in our case).
(3) The encoding/decoding delays are 25ms/10ms on the sender/receiver UEs, respectively.
(4) The software matters: when using FFplay~\cite{ffplay} instead of MPlayer~\cite{mplayer} on the receiver UE to pull and display the video stream, the streaming delay reduces almost by 90ms.

\implications{
While \sys brings modest delay reduction to live streaming scenarios by reducing the network delay, the delay spent on image capture, transcoding, jitter buffer, and even the system-software stack remain the bottleneck to achieve real-time human interaction.
Thus, it is imperative to improve the hardware capacities and the system software design in order to support the niche edge applications.
Even so, live streaming has become a major application type hosted on \sys for its reduced billing cost as we will show in $\S$\ref{sec:cost}.
} 

%% file: tab-app-exp.tex
\begin{table}[t]
\centering
\begin{tabular}{lllll}
 & \textbf{Edge} & \textbf{Cloud-1} & \textbf{Cloud-2} & \textbf{Cloud-3} \\\hline
WiFi & 11.4ms & 16.6ms & 40.9ms & 55.1ms \\\hline
LTE & 22.2ms & 25.6ms & 54.6ms & 63.2ms \\\hline
5G & 18.1ms & 22.8ms  & 49.5ms & 60.8ms \\\hline
\end{tabular}
\caption{\textbf{The RTTs of edge/cloud VMs used for QoE experiments in $\S$\ref{sec:app-performance}}, averaged across different locations.}
\label{tab:app-perf}
\end{table}

%% file: fig-game.tex

\begin{figure}[t]
	\centering
	\begin{minipage}[b]{0.44\textwidth}
		\includegraphics[width=1\textwidth]{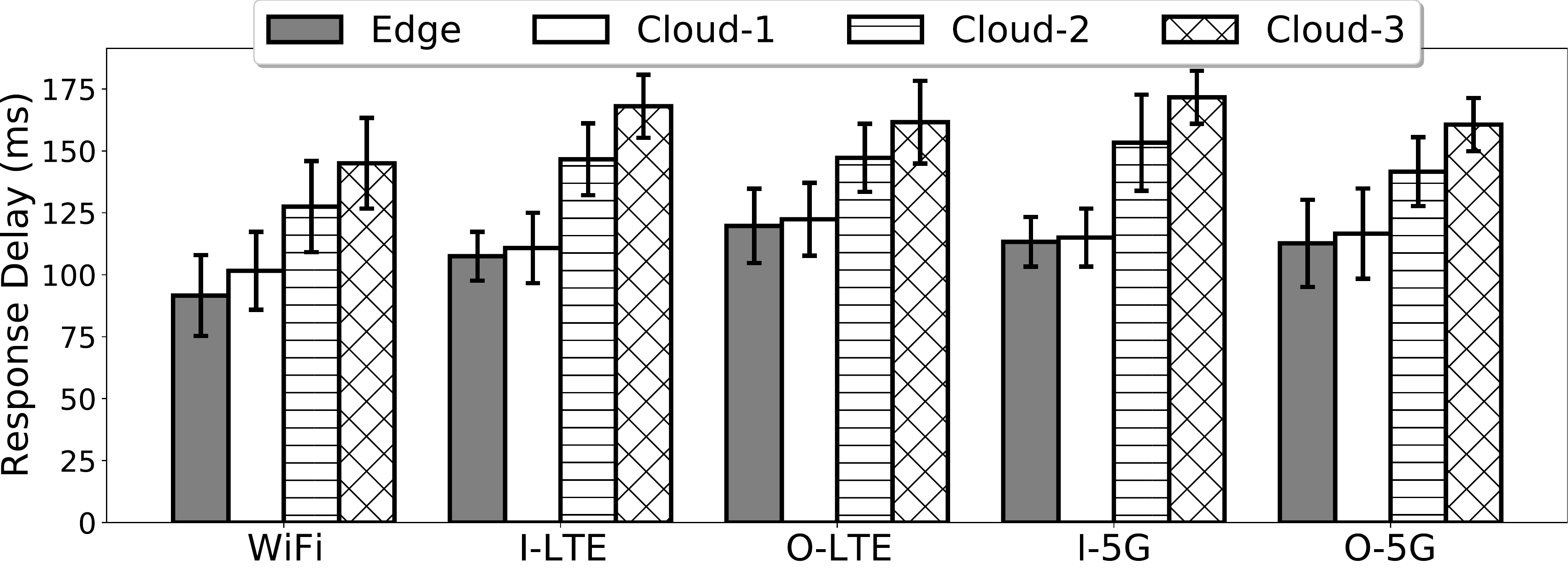}
		\subcaption{Under different network conditions.}
	\end{minipage}

	\centering					
	\begin{minipage}[b]{0.24\textwidth}
		\includegraphics[width=1\textwidth]{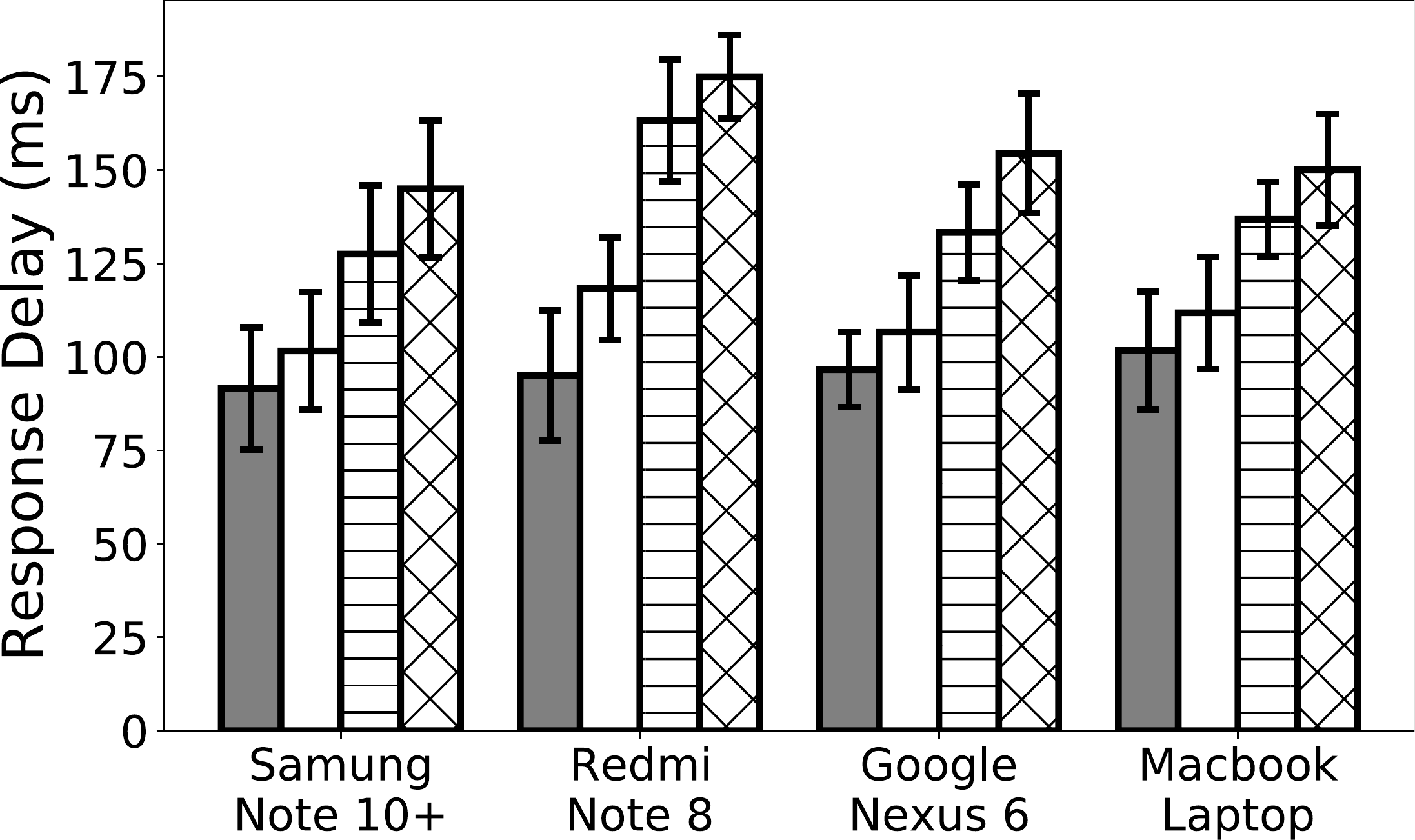}
		\subcaption{With different devices.}
	\end{minipage}
	~
	\begin{minipage}[b]{0.20\textwidth}
		\includegraphics[width=1\textwidth]{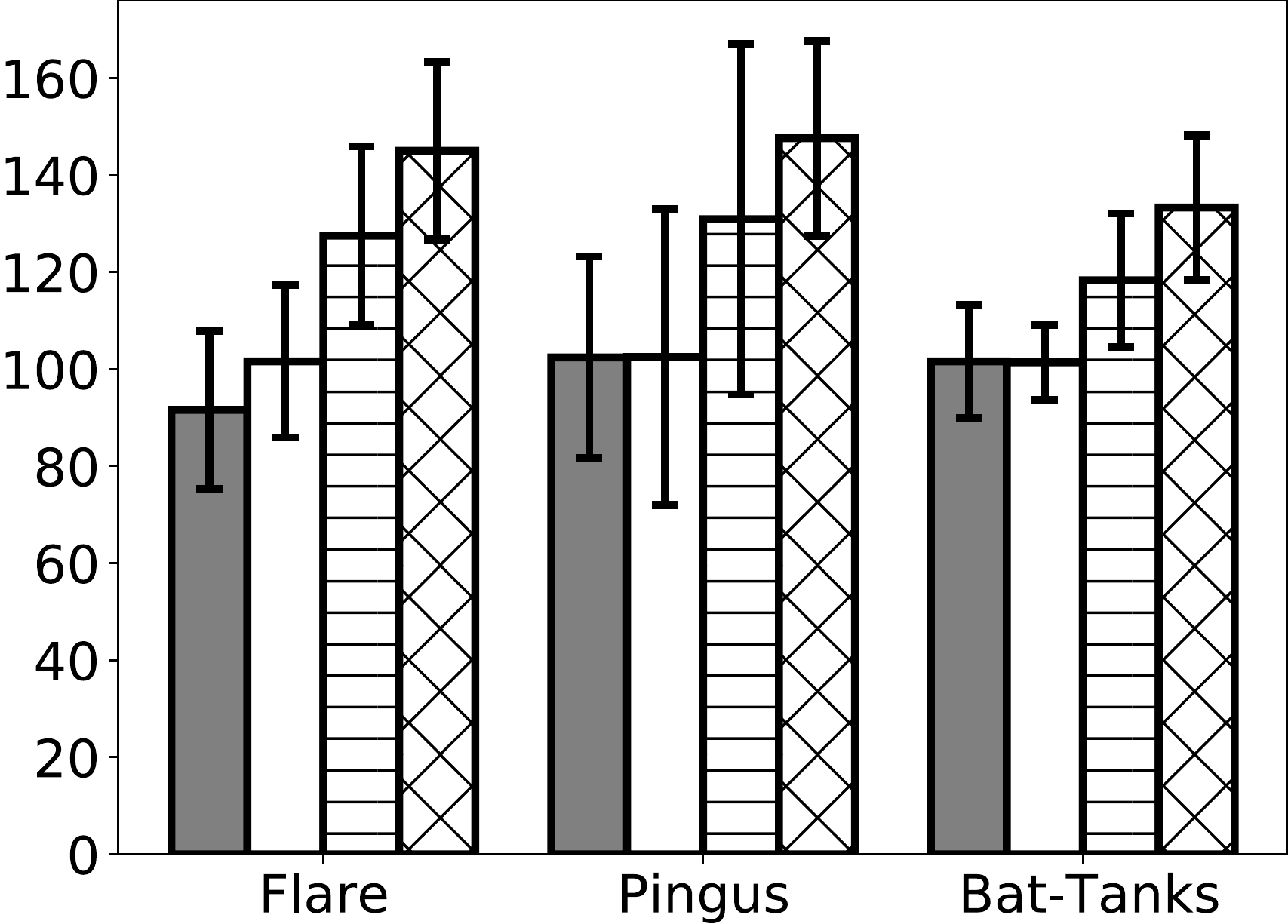}
		\subcaption{Using different games.}
	\end{minipage}

\caption{\textbf{The cloud gaming performance under different settings.}
Default setting: Samsung Note 10+, Game \textit{Flare}, and WiFi.
``I-'': indoor, ``O-'': outdoor.
}
\vspace{-15pt}
\label{fig:cloud-gaming}
\end{figure}

%% file: fig-streaming.tex

\begin{figure}[t]
	\centering
	\begin{minipage}[b]{0.4\textwidth}
		\includegraphics[width=1\textwidth]{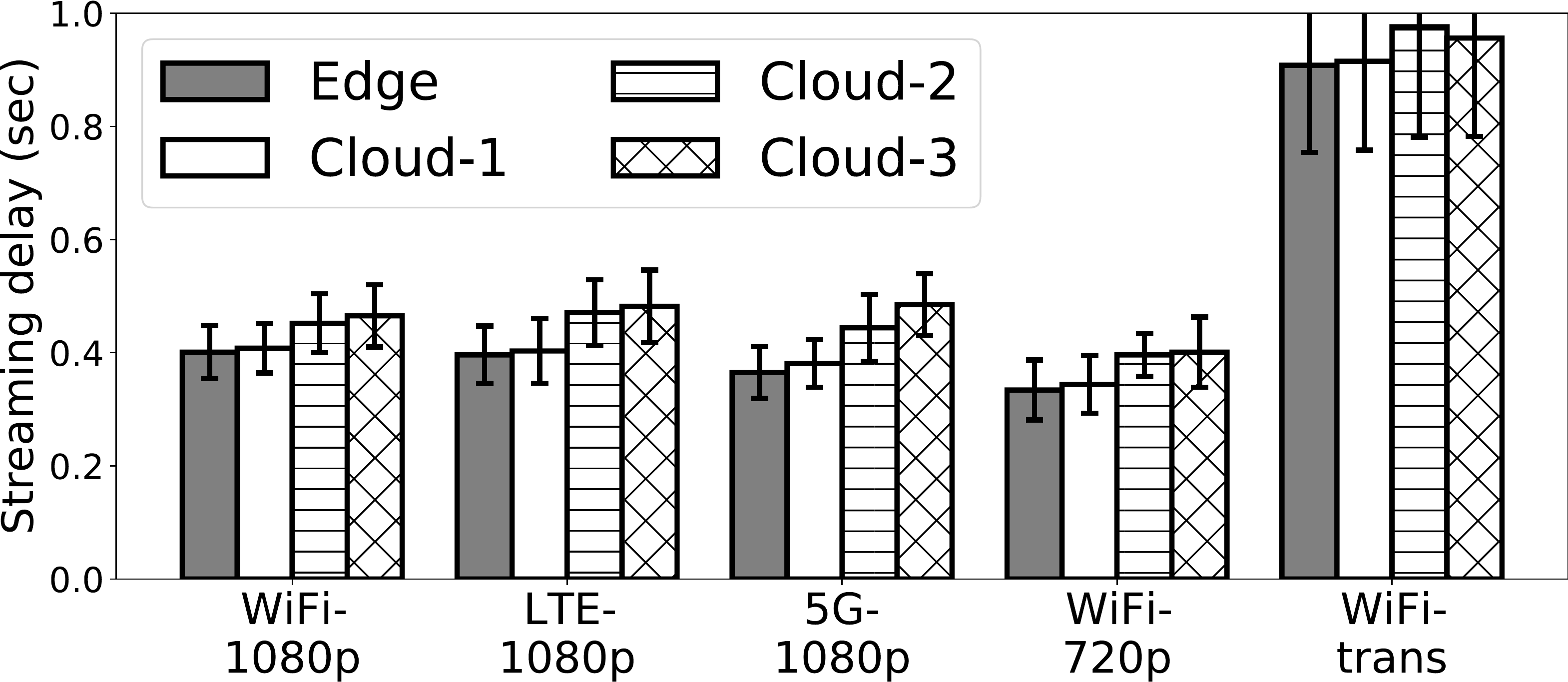}
	\end{minipage}
	
	\caption{\textbf{The live streaming performance on edge/cloud under different experiment conditions}.}
	\vspace{-15pt}
	\label{fig:live-streaming}
\end{figure}

%% file: analysis.tex
\section{Demystifying Edge Workloads}\label{sec:workloads}

In this section, we characterize the workloads based on our collected VM traces ($\S$\ref{sec:data-passive}) and the public \textbf{Azure Dataset} collected from Azure's \textit{entire} VM sets~\cite{cortez2017resource} (2019 version).


%
%
%

\subsection{Applications and VM Subscription}\label{sec:vm-type}
\textbf{Application type}
We investigate the major customers of \sys.
We classify them into different categories, and the most popular ones are:
video live streaming, online education, content delivery, video/audio communication, video surveillance, and cloud gaming.
Most of them have two common characteristics:
(1) Network-intensive: they stream a lot of data, mostly videos. $\S$\ref{sec:cost} will show that edge platforms like \sys are more budget-friendly to the applications with high bandwidth usage.
(2) Delay-critical: user interaction is often involved, either unidirectional or bidirectional. This is because edge services can provide lower and more stable network performance.
In fact, those two factors are the major incentives to decentralize cloud applications to edges.

\input{fig-vm-size}

\textbf{VM size}
We compare the VM sizes, i.e., the amount of resources allocated to each VM on \sys and Azure Cloud.
\revise{Note that \sys provides very similar VM configuration options in terms of CPU and memory to customers as Azure Cloud does.}
Illustrated in Figure~\ref{fig:vm-size},
our key observation is that \sys VMs typically request more resources than Azure VMs.
Overall, the median number of CPU cores and memory requested on \sys and Azure are (8 vs. 1) and (32GBs vs. 4GBs).
In addition, Azure Cloud has many ``low-end'' VMs with only a few CPU cores (90\% VMs with $\leq$4 vCPUs) and relatively little memory (70\% VMs with $\leq$4 GBs),
while \sys's half VMs have more than 8 CPU cores and 16GBs memory.
The median/mean storage size of \sys VMs is 100/650 GBs (not compared as Azure dataset doesn't contain storage information).

The possible reason for \sys VMs subscribing more hardware resources is that \sys's current customers are mostly business-oriented, who need to deploy commercial services or apps that are likely to be delay-critical, while Azure also serves individuals (e.g., researchers, educators) who only need very few resources per VM to complete their jobs.

\implications{
Large VM size \revise{on \sys-like edge platforms} may cause severe resource fragmentation, i.e., the bin-packing problem~\cite{conf/asplos/DelimitrouK14,meisner2011powernap}, hindering a high sale ratio for each server as we will show next.
To mitigate such fragmentation, techniques like dynamic VM migration~\cite{mishra2012dynamic} and resource disaggregation~\cite{legoos} may help.
}

\input{fig-vm-deployment}

\textbf{VM numbers per app}
As shown in Figure~\ref{fig:deployment}, customers tend to deploy \textit{slightly more} VMs on \sys than Azure.
For instance, more than 9.6\% apps on \sys deploy at least 50 VMs, while on Azure only 6.1\% of apps deploy that many.
The largest edge app is a CDN application which comprises of almost 1,000 VMs.
There exist two possible reasons:
(1) Apps deployed on \sys are more likely to be delay-sensitive than Azure, requiring a larger number of geo-distributed VMs to guarantee low delay and high reliability.
(2) As aforementioned, Azure serves more small-scale businesses or individuals who only need very few VMs to operate.

\implications{
Compared to clouds, managing and scheduling a large number of geo-distributed edge VMs is more challenging. 
First, traditional tools like Kubernetes~\cite{kubernetes} may not suffice in maintaining such VM orchestration~\cite{kubernetes-err}. It motivates us to improve or re-architect those tools.
Second, edge customers also need more effort to schedule end-user traffic to the optimal VM in a fine-grained way.
$\S$\ref{sec:resource-allocate} will show that current edge customers often fail to make a good scheduling decision.
}

\textbf{Servers/\nodes sales rate}
We also summarize the resource sales rate on \sys (figure not shown), defined as the percentage of CPU/memory resources sold out to customers per \node or server.
We have two key observations:
(1) The sales rate is highly skewed across servers and \nodes.
For example, the 95th-percentile CPU sales rate across \nodes is about 5$\times$ higher than the 5th-percentile.
Such skewness stems from that, in edge computing, server resource demand highly depends on the geolocations.
(2) Compared to memory, CPU cores are more likely to be saturated: the median sales rate of CPU is almost 2$\times$ of the memory.

\implications{
The skewed sales rate across \nodes can guide \revise{\sys providers} to locate the regions with higher business returns for future investment. As edge apps increasing, it's also important to invest more in those ``hot spots'' to ensure good availability and elasticity of computation resources.
}

\subsection{Overall Resource Usage}\label{sec:vm-usage}

\input{fig-cpu-usage}

\textbf{Overall CPU usage}
Figure~\ref{fig:cpu-usage}(a) illustrates the per-VM CPU usage on edge/cloud as the Cumulative Distribution Function (CDF) of the average CPU utilization, and the
CDF of the 95th-percentile of the maximum CPU utilization (P95 Max).
The key observation is that, either by mean or P95 Max, VM CPUs on \sys are much less utilized than Azure Cloud.
For example, 74\% VMs on \sys have less than 10\% CPU utilization on average, while on Azure Cloud only 47\% VMs have less than 10\% utilization.
It indicates that, either unconsciously or purposely, edge customers tend to over-provision the hardware resources for VMs, which also echoes our analysis in $\S$\ref{sec:vm-type} that edge VMs often subscribe more resources than on cloud.
Diving deeper, such resource over-provision may be due to two reasons:
(1) Edge apps are more likely to be delay-critical, so that more resources are needed to deliver a good quality of service.
(2) It is difficult for edge customers to understand the resource demand at different locations \revise{due to the high density of edge server deployment and the temporal dynamics of user requests as will be discussed below.}
There, they tend to be conservative when provisioning them.

\textbf{CPU usage variance across time}
We also investigate the across-time resource usage variance of edge/cloud VMs, as indicated by the coefficient of variation (CV = std/mean) of CPU usage.
As illustrated in Figure~\ref{fig:cpu-usage}(b), edge VMs exhibit more usage variance across time than cloud (median: 0.48 vs. 0.24).
The reason is that apps deployed on edge platforms are more likely to be interactive ($\S$\ref{sec:vm-type}) so that the usage highly depends on human activities.

\implications{
The relatively low but highly skewed CPU usage challenges the \revise{\sys's VM management}.
To better utilize the CPU resources, \sys may borrow existing techniques from cloud computing research, e.g., smart VM placement algorithms based on VM resource usage prediction~\cite{xvmotion,clark2005live,hadary2020protean}.
An alternative approach is to employ more elastic computing forms, e.g., containers, together with IaaS VMs on the same server.
$\S$\ref{sec:discuss} will further discuss the opportunities and challenges.
}

\subsection{Resource Load Balance}\label{sec:resource-allocate}

\input{fig-cpu-var-nc}

\input{fig-var-usr}

Resource usage balance (or load balance) is critical to the application QoE on multi-tenant cloud platforms~\cite{schad2010runtime,ananta,duet}.
\revise{
Such importance is further amplified on edge platform that hosts delay-critical apps and needs to provide high uptime SLA.
For example, excessively high CPU usage will cause compute tasks to be delayed and high bandwidth usage may cause traffic congestion and long network delay.
}
Therefore, we investigate the load balance of \sys and Azure, from the perspectives of both physical servers/\nodes and apps.

\textbf{Load balance from servers/\nodes perspective}
\footnote{Since the Azure dataset doesn't contain the VM placement information, we cannot make the comparison here.}
We find the resource usage is highly skewed.
Figure~\ref{fig:resource-balance} shows a case of 11 \nodes from the same Province in China and the servers in one of the selected \nodes.
As observed, across servers in the same \node the bandwidth usage gap can be up to 19.8$\times$.
The gap is even more significant across \nodes, i.e., 8.7$\times$ for P95 Max CPU usage and 731$\times$ for bandwidth usage.

\input{fig-bw-var-week}
Note that load balance is one of the major targets in \sys's VM placement strategy ($\S$\ref{sec:ens}).
The above unbalanced load can be accounted to three main reasons.
(1)
Even for a given VM, its resource usage can change dramatically over time.
Figure~\ref{fig:bw_var_week} shows an example of 4 random VMs' usage within three months.
For 2 VMs among them (``VM-1'' and ``VM-2''), the weekly-averaged bandwidth usage varies in a dramatic and unpredictable way.
\revise{In the extreme case, a VM's (black line) bandwidth usage goes down from almost 12Gbps to 4Gbps, and then 0.2Gbps within three consecutive weeks (6th-8th), and then goes up back to 4Gbps in the 12th week.}
(2) As \sys is still evolving rapidly, new \nodes are added to \sys frequently.
This also explains why the resource usage skewness is more severe across \nodes than servers.
With the arrival of both \nodes and VM subscriptions, it becomes difficult to balance the resource usage.
(3) The strategies of VM allocation and end-user request scheduling are made by edge providers and customers independently. Such a separation hinders load balancing.

\implications{
Given the observed dynamics and complexity of VM resource usage, we found it's extremely difficult to design an effective \textit{static} resource allocation strategy.
Instead, we envision that \textit{dynamic} VM migration~\cite{xvmotion,clark2005live} can better balance the across-server resource usage.
Perhaps a more fundamental approach is changing the way of resource allocation from VM-based to more elastic ones (e.g., FaaS or serverless~\cite{jonas2019cloud,aws-serverless}).
}

\textbf{Load balance from app perspective}
We further investigate the resource usage of VMs from the same app.
The results are illustrated in Figure~\ref{fig:var_usr}, where
subplot (a) shows how unbalanced the VM loads from the same app are.
Our major observation is that there are much more apps with highly unbalanced cross-VM usage on edge than cloud.
For instance, 16.3\% of edge apps have more than 50$\times$ cross-VM usage gap (defined as the 95th-percentile divided by the 5th-percentile of the mean CPU usage of all the VMs that an app uses) on \sys, while on Azure only 0.1\% of apps have that large unbalanced CPU usage.
Figure~\ref{fig:var_usr}(b) zooms into one app and shows its 11 VMs' CPU utilization in a day (one curve corresponds to one VM).
As observed, there's one VM running at very high load, e.g., e.g., for more than 33\% of the time, the CPU utilization is higher than a typical safe threshold of 80\%.
In contrast, some other VMs' CPU utilization is constantly below 30\%.

\implications{
Given the importance of load balance, however, our results demonstrate that current edge apps deployed on \sys often fail to deliver this goal.
Indeed, achieving load balance on edge platforms is difficult, because the VMs are geo-distributed and their resource usage patterns may change over time as aforementioned (Figure~\ref{fig:bw_var_week}).
To handle this challenge, we can resort to:
(1) Dynamically adapting the resources of each VM to what's needed.
This approach, however, requires rebooting the VM that can take tens of seconds or even minutes, as current edge/cloud platforms don't support hot resource scaling.
(2) Load-aware traffic scheduling from end users to VMs considering the current loads on each available VM.
This is similar to global server loading balancing (GSLB) commonly used in web traffic management and application data delivery~\cite{dilley2002globally}.
However, edge customers face unique challenges in load balancing because inter-\node request scheduling may increase the user-perceived network delay.
Even so, we believe a load balancer is useful in edge platforms as the network delay between nearby edge \nodes is already small ($\S$\ref{sec:network-delay}).
}

\input{prediction}

\input{cost}

%% file: fig-vm-size.tex
\begin{figure}[t]
	\centering					
	\begin{minipage}[b]{0.23\textwidth}
		\includegraphics[width=1\textwidth]{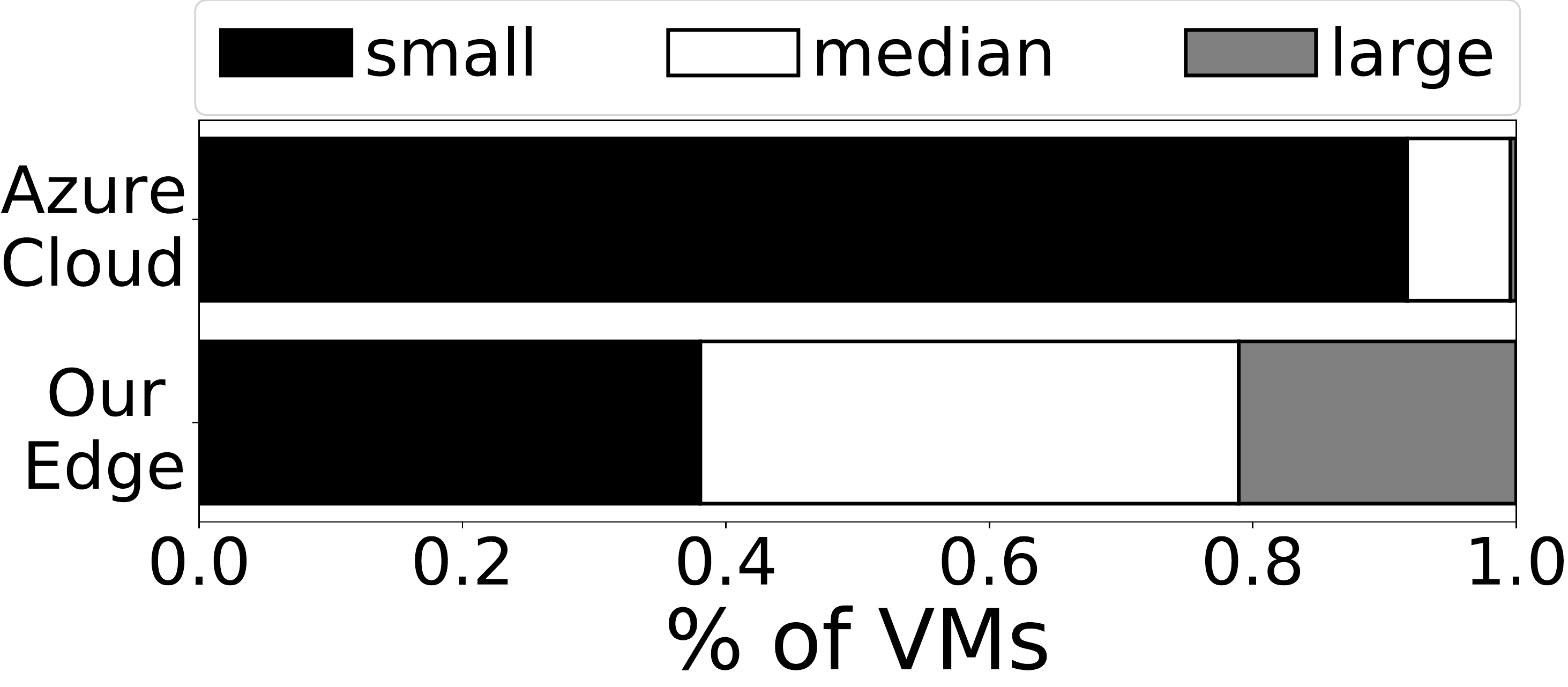}
		\subcaption{\# of vCPUs per VM}
	\end{minipage}	
	~
	\begin{minipage}[b]{0.23\textwidth}
		\includegraphics[width=1\textwidth]{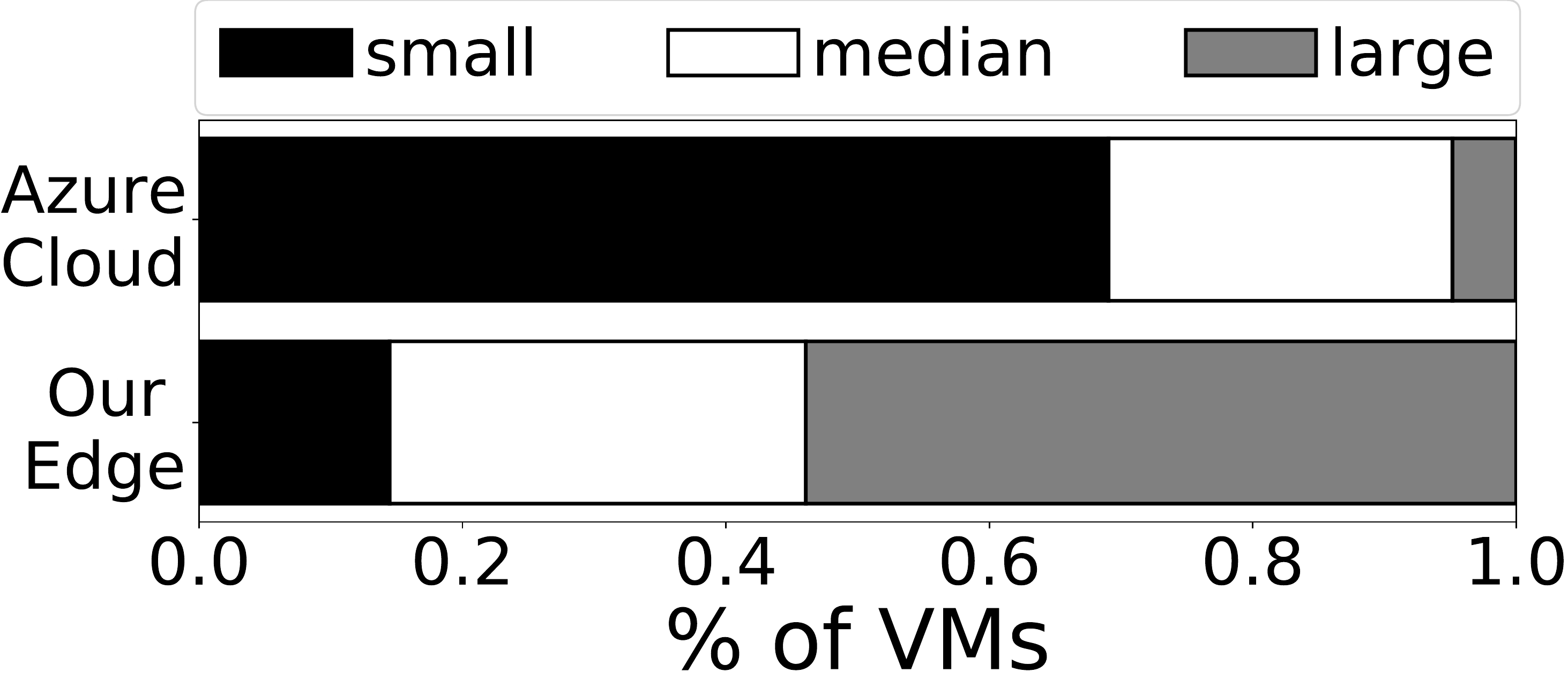}
		\subcaption{Memory size per VM}
	\end{minipage}
	
	\caption{\textbf{\sys VMs are larger than Azure.}
		``small/median/large'': $\le$4 / 5--16 / $>$16 CPU core or GBs memory.
	}
	\vspace{-15pt}
	\label{fig:vm-size}
\end{figure}

%% file: fig-vm-deployment.tex


\begin{wrapfigure}{r}{0.24\textwidth}
	\centering
	\includegraphics[width=.23\textwidth]{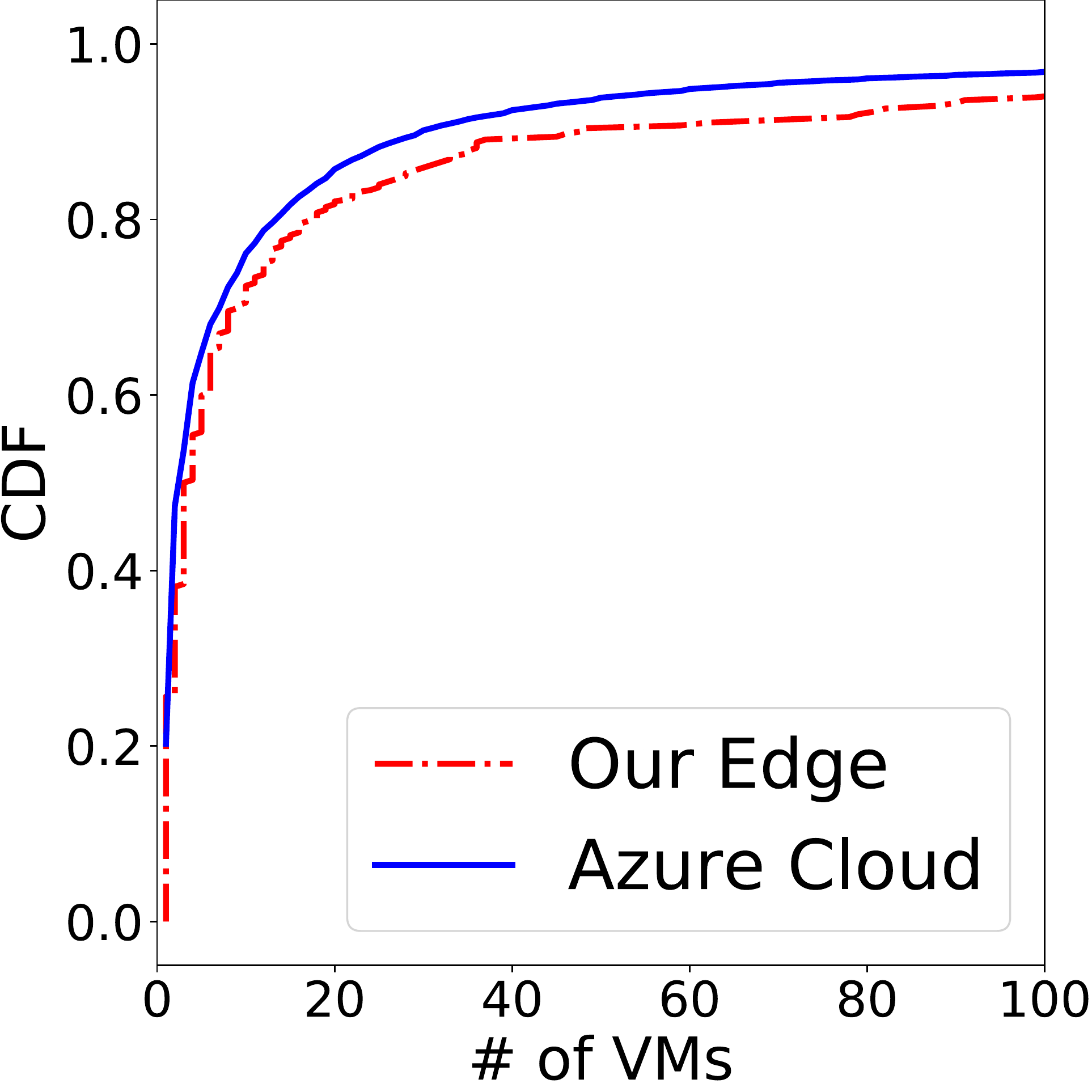}
	\caption{Per-app VM num.}
	\label{fig:deployment}
	\vspace{-10pt}
\end{wrapfigure}

%% file: fig-cpu-usage.tex

\begin{figure}[t]
	\centering					
	\begin{minipage}[b]{0.23\textwidth}
		\includegraphics[width=1\textwidth]{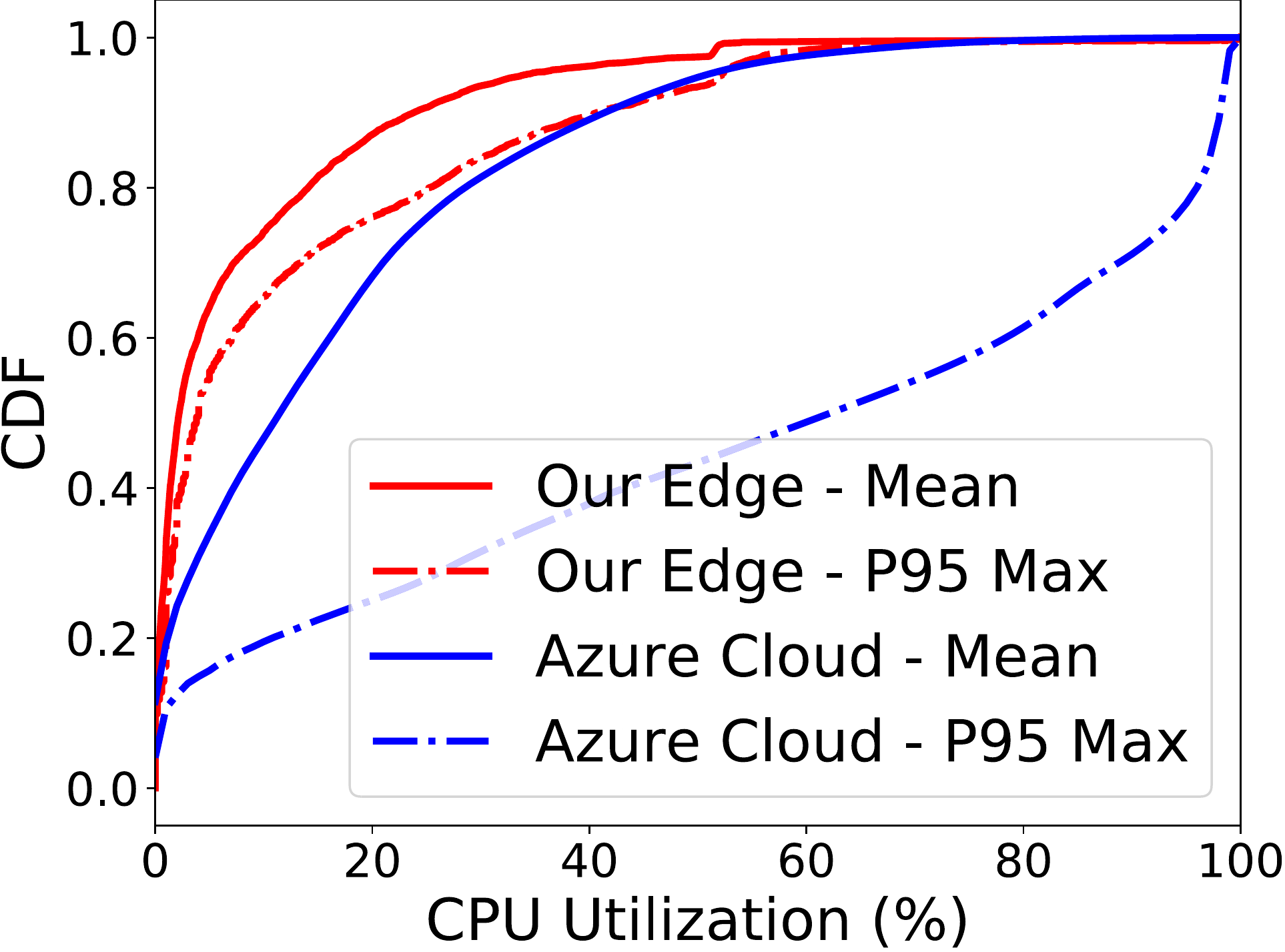}
		\subcaption{Mean/Max CPU util.}
		\label{fig:cpu-usage-overall}
	\end{minipage}	
	~		
	\begin{minipage}[b]{0.23\textwidth}
		\includegraphics[width=1\textwidth]{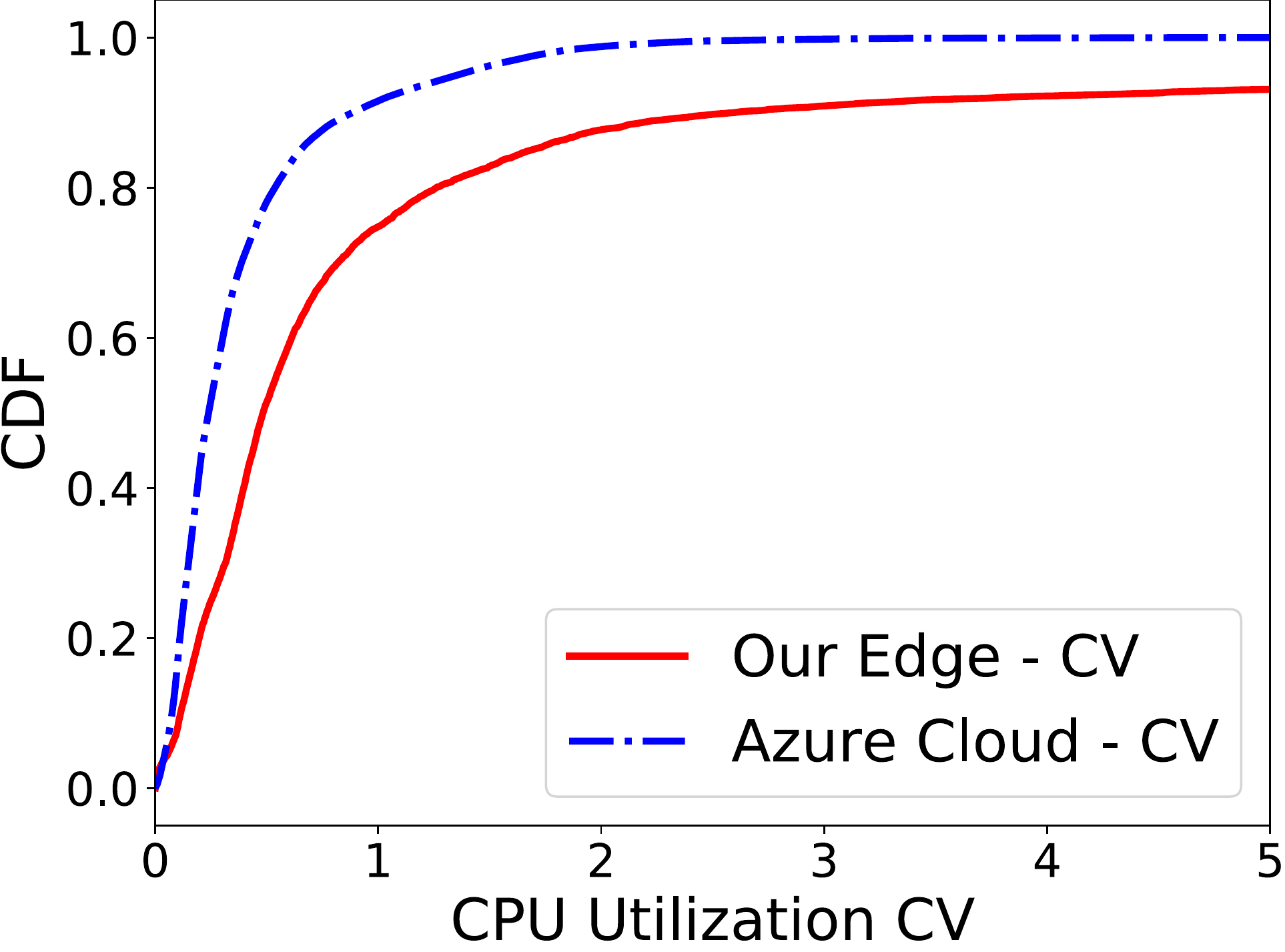}
		\subcaption{CPU util. variance}
		\label{fig:cpu-usage-var}
	\end{minipage}
	
	\caption{\textbf{CPU utilization on \sys is lower but more variant than Azure.}
		(a): the overall CPU utilization; (b): CPU utilization variance across time (CV).}
	\vspace{-15pt}
	\label{fig:cpu-usage}
\end{figure}

%% file: fig-cpu-var-nc.tex
\begin{figure}[t]
	\centering				
	\begin{minipage}[b]{0.22\textwidth}
	\includegraphics[width=1\textwidth]{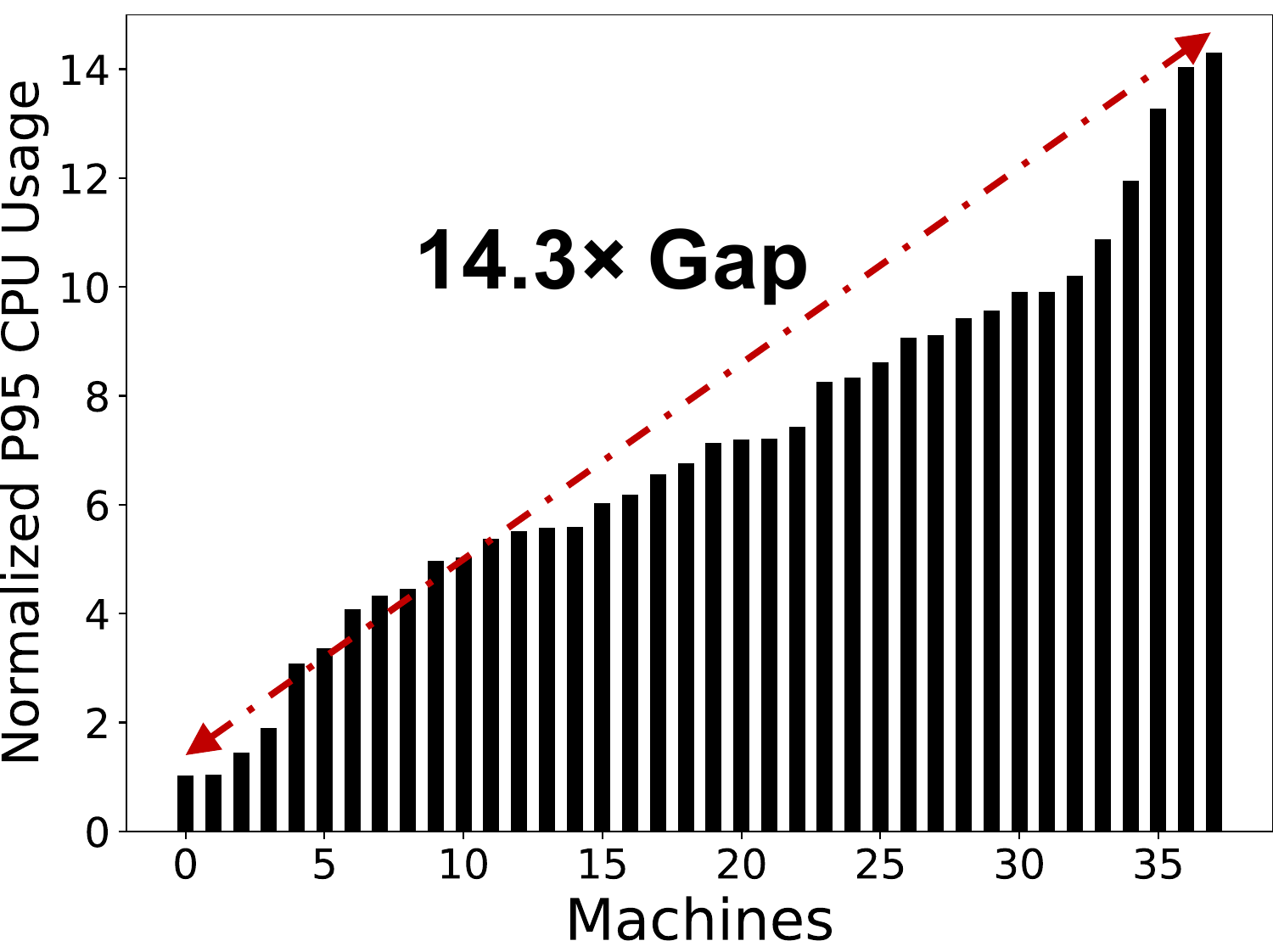}
	\subcaption{Cross-server CPU usage}
	\label{fig:cpu-balance-nc}
	\end{minipage}	
	~		
	\begin{minipage}[b]{0.22\textwidth}
	\includegraphics[width=1\textwidth]{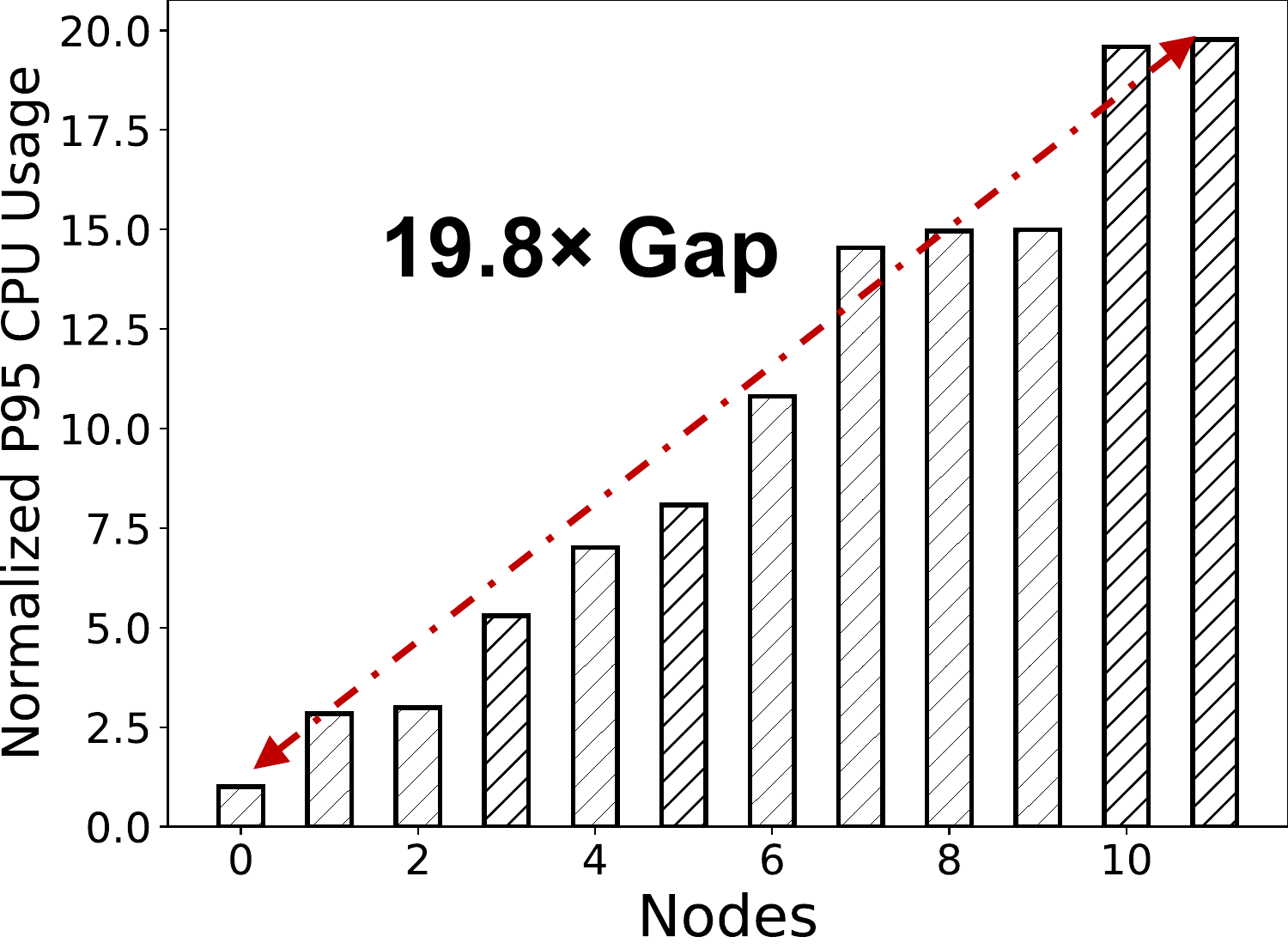}
	\subcaption{Cross-site CPU usage}
	\label{fig:cpu-balance-node}
	\end{minipage}
	\begin{minipage}[b]{0.22\textwidth}
		\includegraphics[width=1\textwidth]{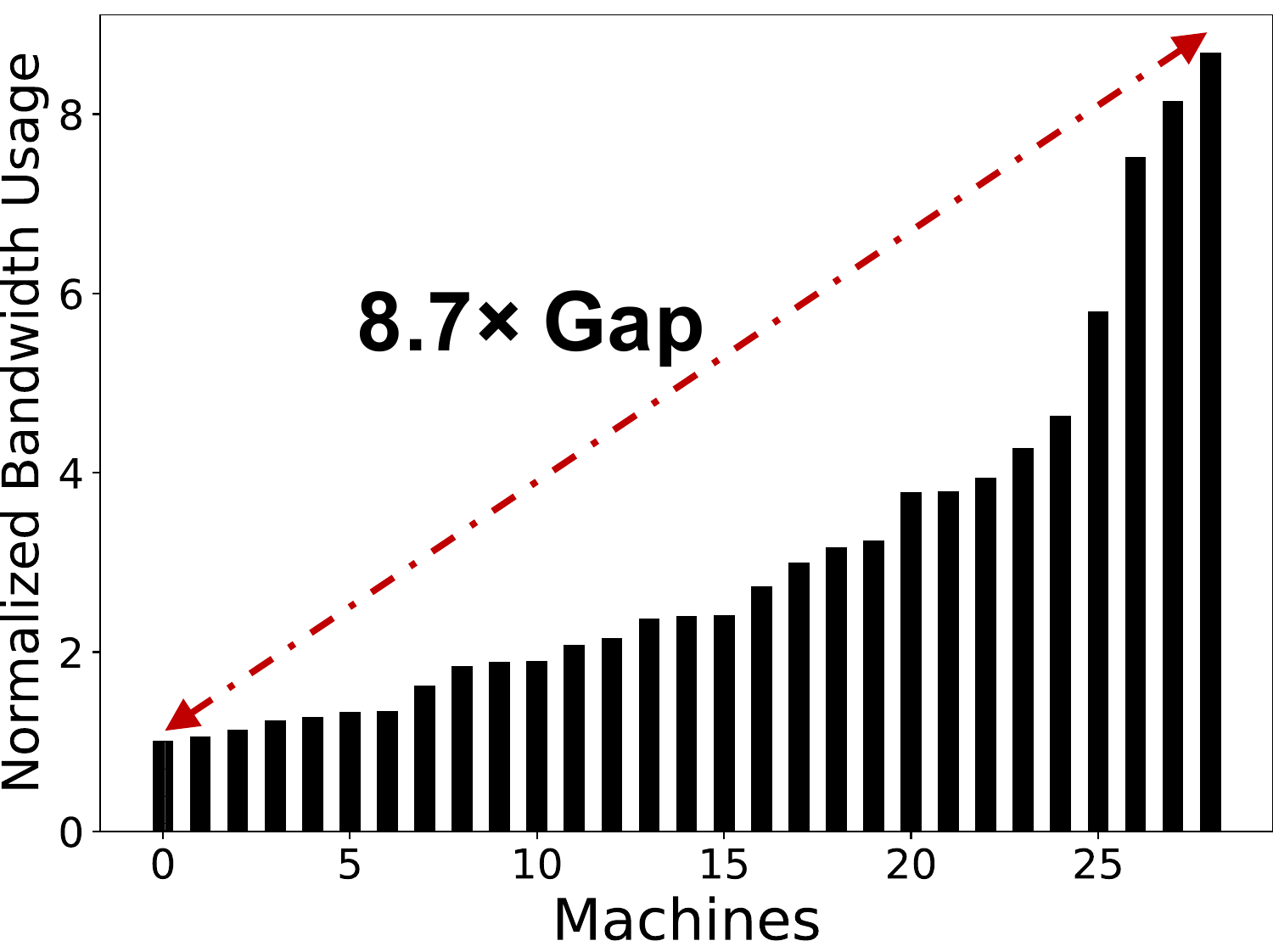}
		\subcaption{Cross-server NET usage}
		\label{fig:bw-balance-nc}
	\end{minipage}	
	~		
	\begin{minipage}[b]{0.22\textwidth}
		\includegraphics[width=1\textwidth]{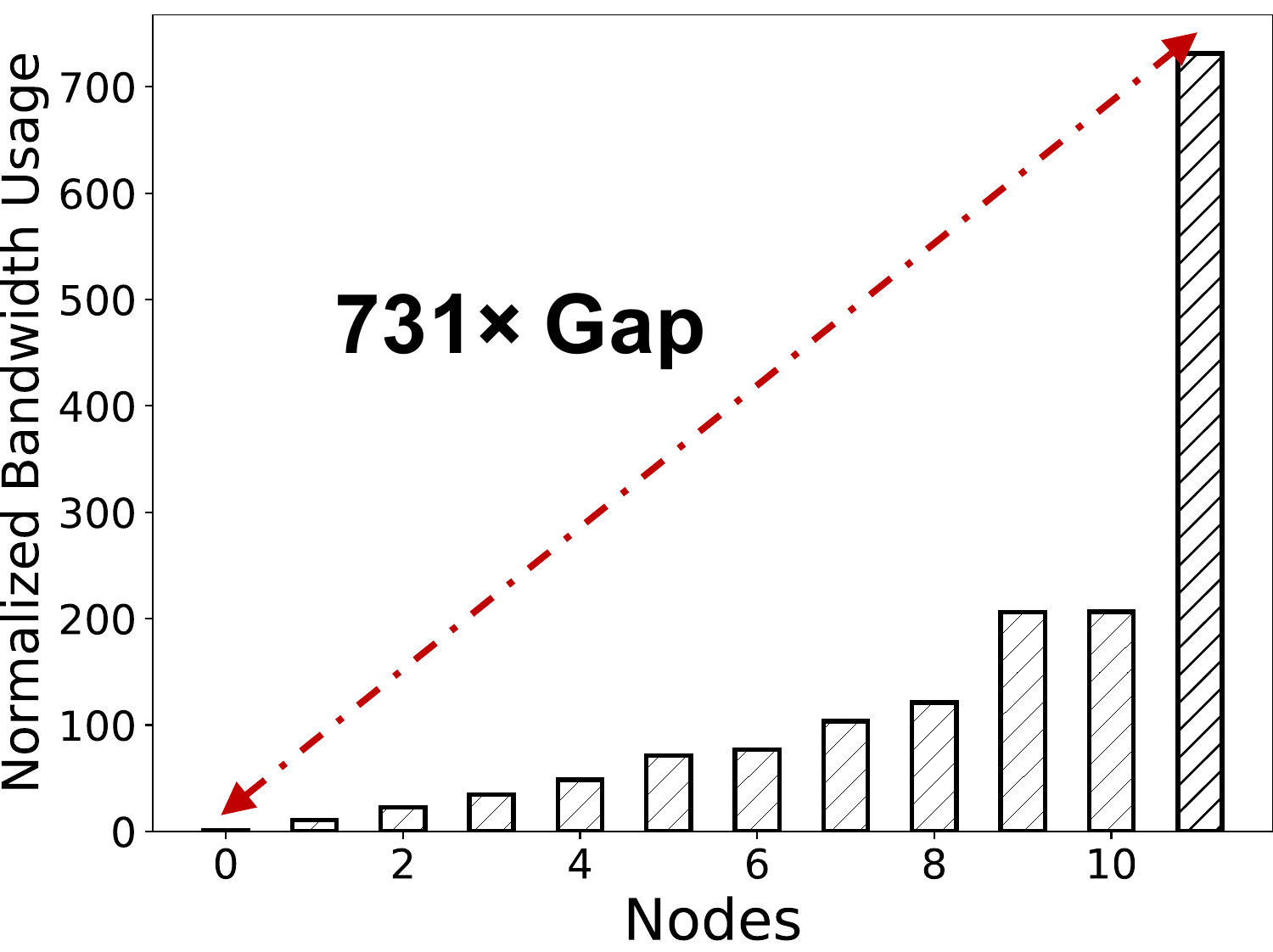}
		\subcaption{Cross-site NET usage}
		\label{fig:bw-balance-node}
	\end{minipage}
	
	\caption{\textbf{The resource usage across machines/\nodes is highly unbalanced.}
		All \nodes are randomly sampled from Guangdong Province, and the machines are from a random \node.
		For (a)/(b): a machine's CPU usage is calculated as the weighted (by requested cores) CPU usage of all its hosted VMs, and a \node's CPU usage is averaged across all its machines.
		For (c)/(d): the bandwidth usage of a machine/\node is summed across all the VMs hosted in the machine/\node.
		For each figure: all numbers are normalized to the smallest one.}
	\vspace{-10pt}
	\label{fig:resource-balance}
\end{figure}

%% file: fig-var-usr.tex

\begin{figure}[t]
	\centering					
	\begin{minipage}[b]{0.23\textwidth}
		\includegraphics[width=1\textwidth]{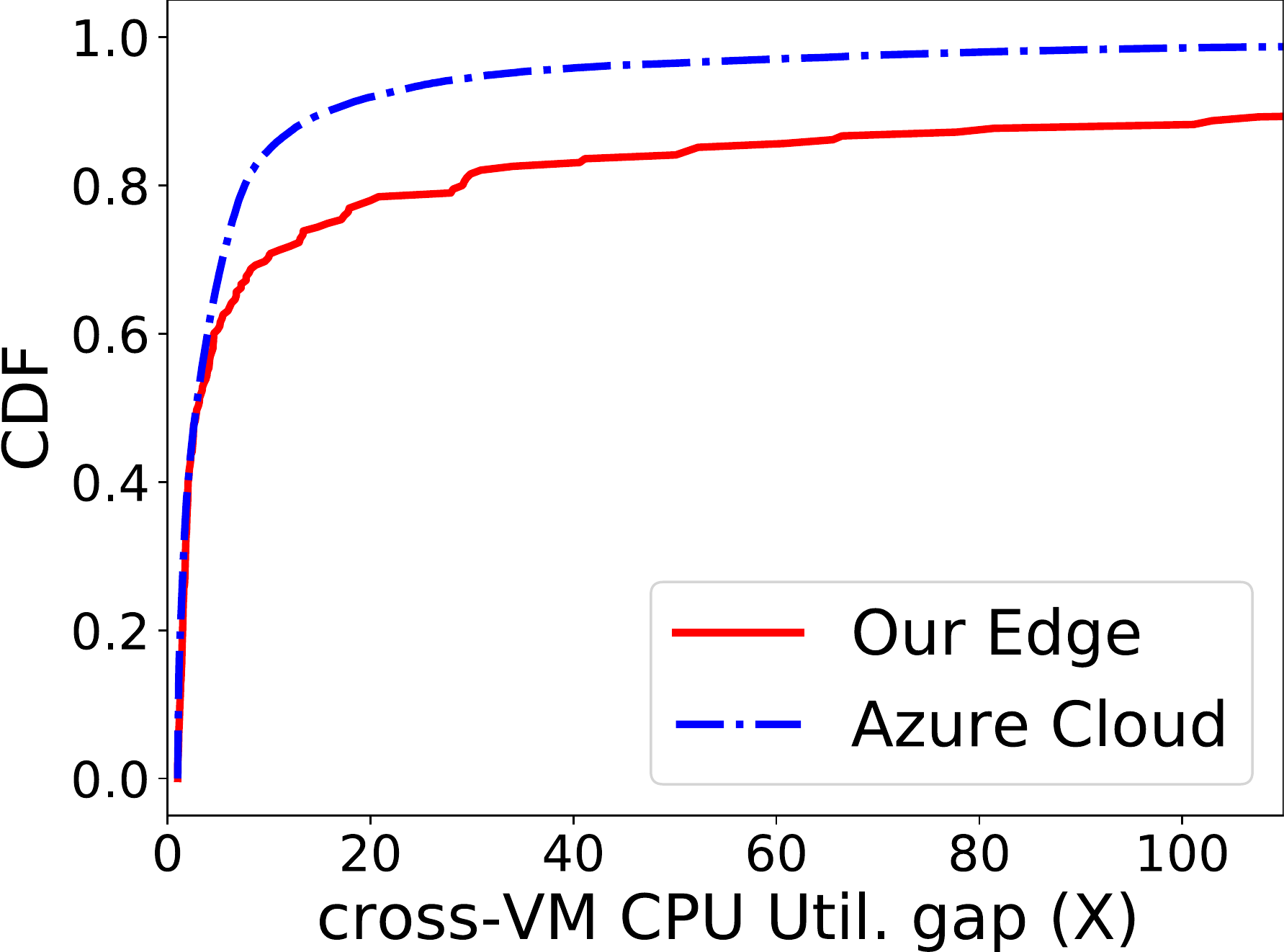}
		\subcaption{CPU usage gap between the same app's VMs.}
	\end{minipage}	
	~		
	\begin{minipage}[b]{0.23\textwidth}
		\includegraphics[width=1\textwidth]{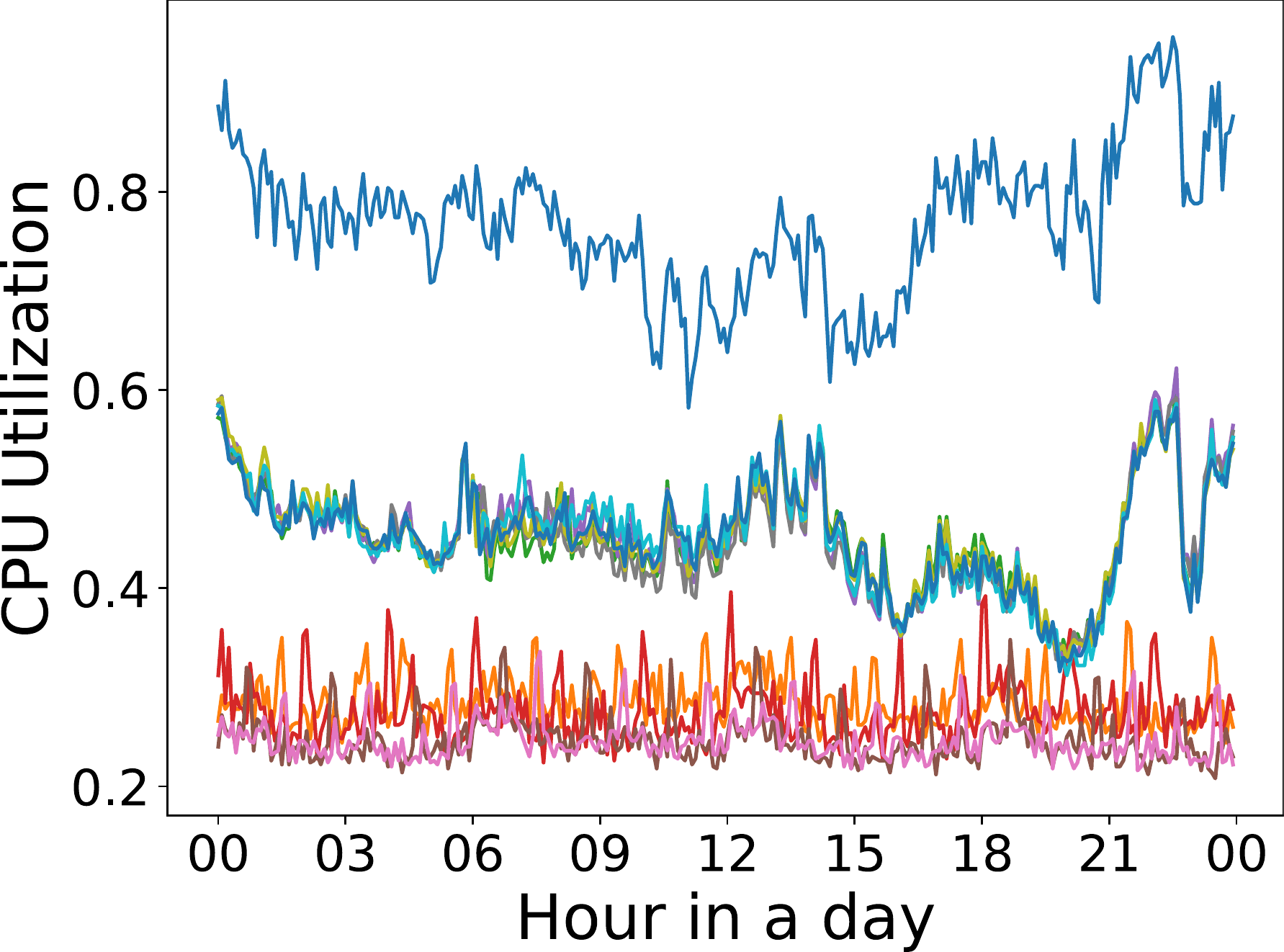}
		\subcaption{CPU usage of 11 VMs from the same edge app.}
	\end{minipage}
	
	\caption{\textbf{The CPU usage of the same app's VMs is highly unbalanced.}
	In (a), the usage gap of each app is measured as the 95th-percentile divided by the 5th-percentile of the mean CPU usage of all its VMs.
	}
	\vspace{-10pt}
	\label{fig:var_usr}
\end{figure}

%% file: fig-bw-var-week.tex
\begin{wrapfigure}{r}{0.24\textwidth}
	\centering
	\includegraphics[width=.23\textwidth]{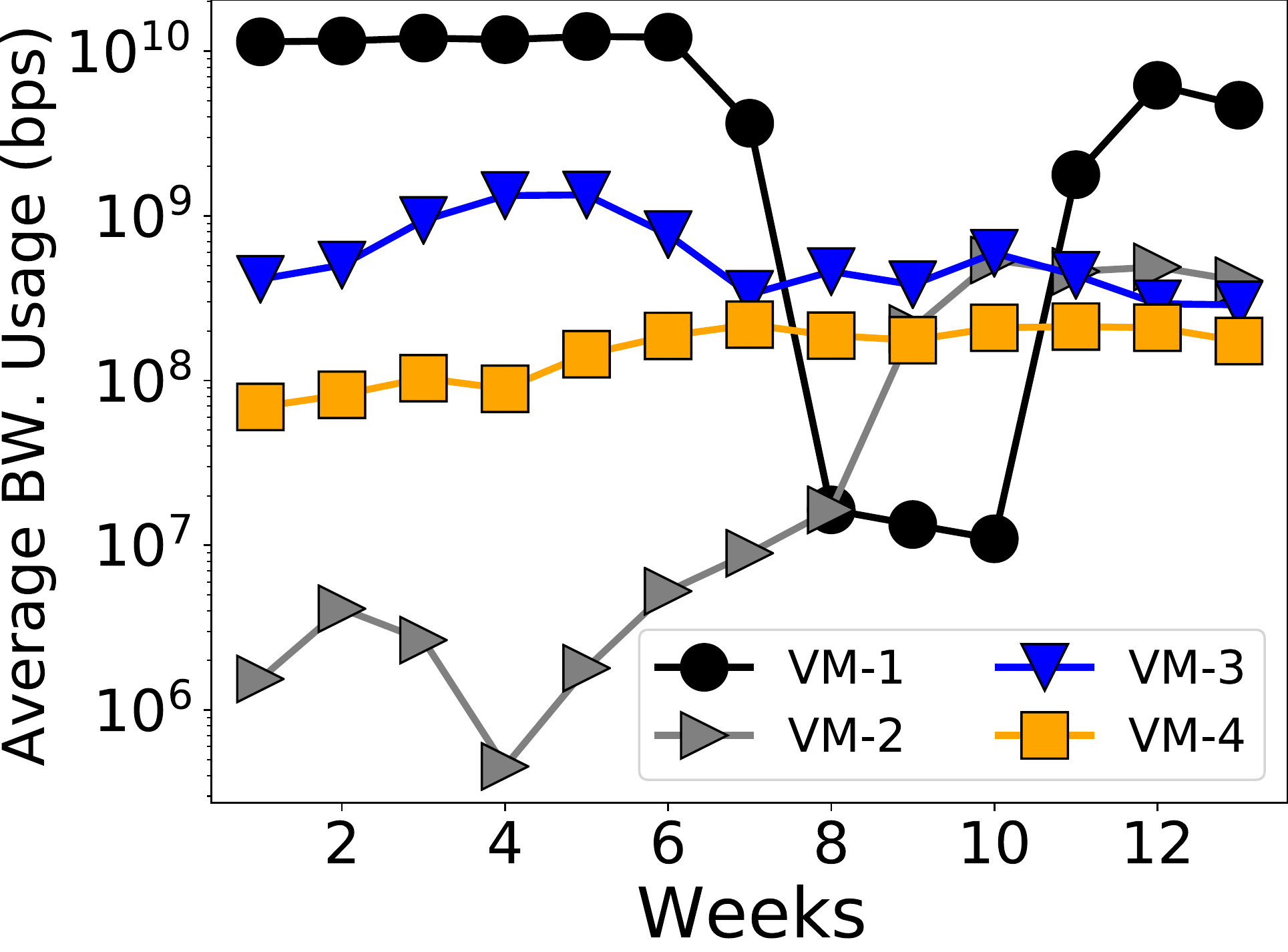}
	\caption{\textbf{The bandwidth usage of VMs may vary significantly across time}.
	}
	\vspace{-10pt}
	\label{fig:bw_var_week}
\end{wrapfigure}

%% file: prediction.tex
\subsection{VM Usage Prediction}\label{sec:pred}

VM usage prediction is a critical feature in data center management~\cite{cortez2017resource,gong2010press,calheiros2014workload}.
In this section, we compare the difficulties of VM usage prediction for edge and cloud.
To be fair, we use 1-month data from both Azure and \sys, with each VM's data split to ML training (3 weeks) and testing (1 week).
The task is to predict the max/mean CPU usage of the next half-hour window based on the historical data.
We try two algorithms, Holt-Winters~\cite{holt-winters} and LSTM~\cite{hochreiter1997long}, which are commonly used for workloads prediction~\cite{conf/sigmod/HigginsonDAPE20}.
The LSTM model has 1 layer and 24 units (2496 weights).
The two models are trained and tested on each separated VM for predicting maximal and mean usage, respectively.
We use root mean square error (RMSE) as the metric to obtain the prediction accuracy.

\input{fig-pred-usage}

Figure~\ref{fig:pred-usage} shows the prediction accuracy of max and mean CPU usage.
Both algorithms achieve higher accuracy on the edge workloads.
For example, the Holt-Winters algorithm achieves a 2.4\% error in predicting the maximal CPU usage on \sys's workloads, much lower than on Azure Cloud (8.5\%).
To predict the mean CPU usage, \sys also has higher accuracy than Azure Cloud but the difference is smaller.
The reason is that the prediction accuracy is already high enough for both workloads (median error $\le$2\%).

The notable difference shown in Figure~\ref{fig:pred-usage} comes from the disparate characteristics of edge and cloud.
We dig into the reason by calculating the VMs' seasonality~\cite{journals/datamine/WangSH06}, an indicator of the strength of the usage patterns across time.
It shows that edge VMs experience stronger seasonality (mean: 0.42) than cloud VMs (mean: 0.26).
Apparently, with stronger seasonality, ML algorithms can better predict the usage based on historical data.
The high seasonality is possibly attributed to the fact that more services deployed on edges follow end users' daily activities.

\implications{
With stronger seasonality and better predictability compared to cloud VMs, edge VMs offer a good opportunity for more fine-grained, smarter resource management.
For example, knowing the future CPU usage can guide VM allocation and migration, thus help avoid server malfunction or even crash induced by CPU overload or network congestion.
}

%% file: fig-pred-usage.tex

\begin{figure}[t]
	\centering					
	\begin{minipage}[b]{0.23\textwidth}
		\includegraphics[width=1\textwidth]{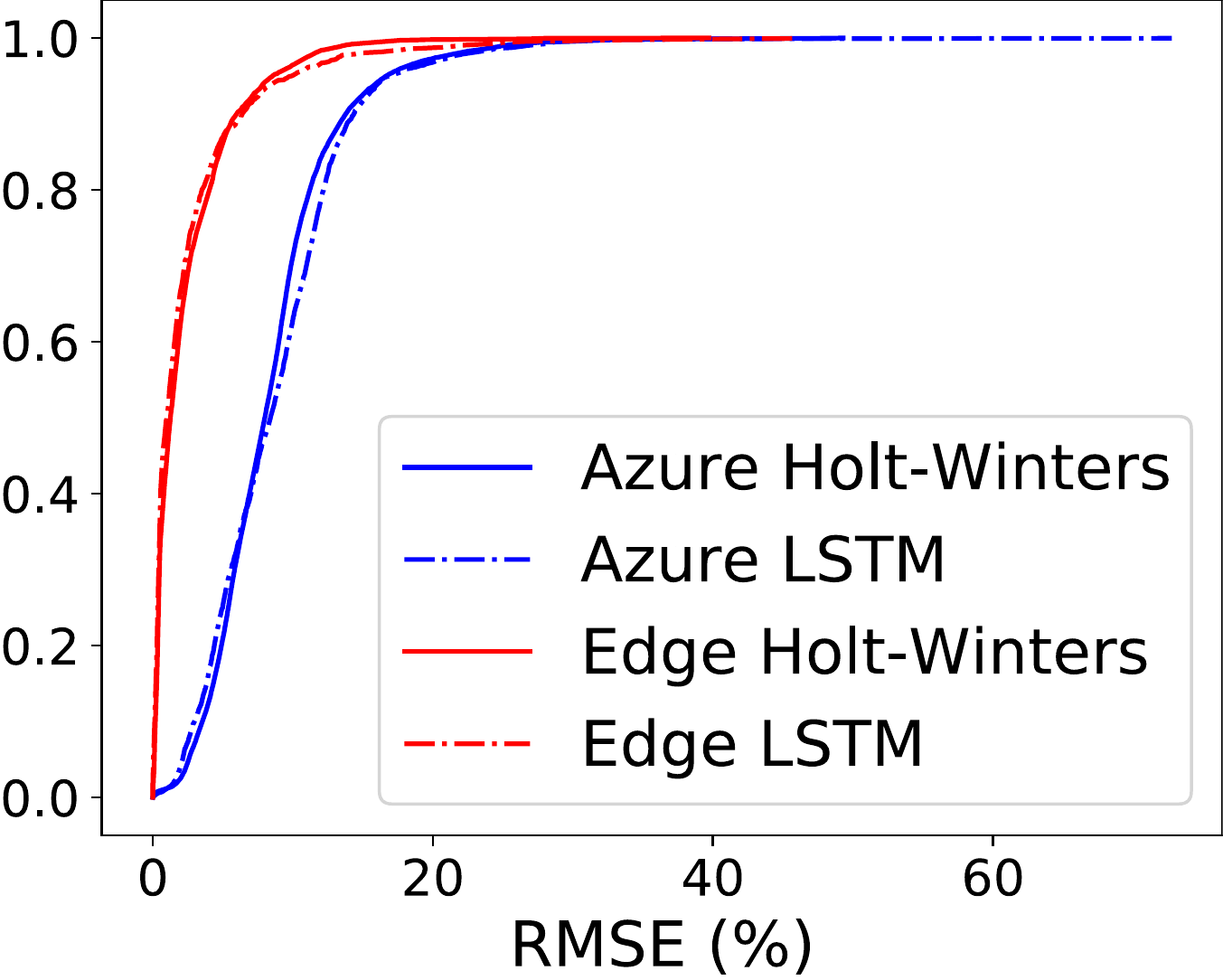}
		\subcaption{Maximal CPU prediction}
		\label{fig:pred-usage-max}
	\end{minipage}	
	~		
	\begin{minipage}[b]{0.23\textwidth}
		\includegraphics[width=1\textwidth]{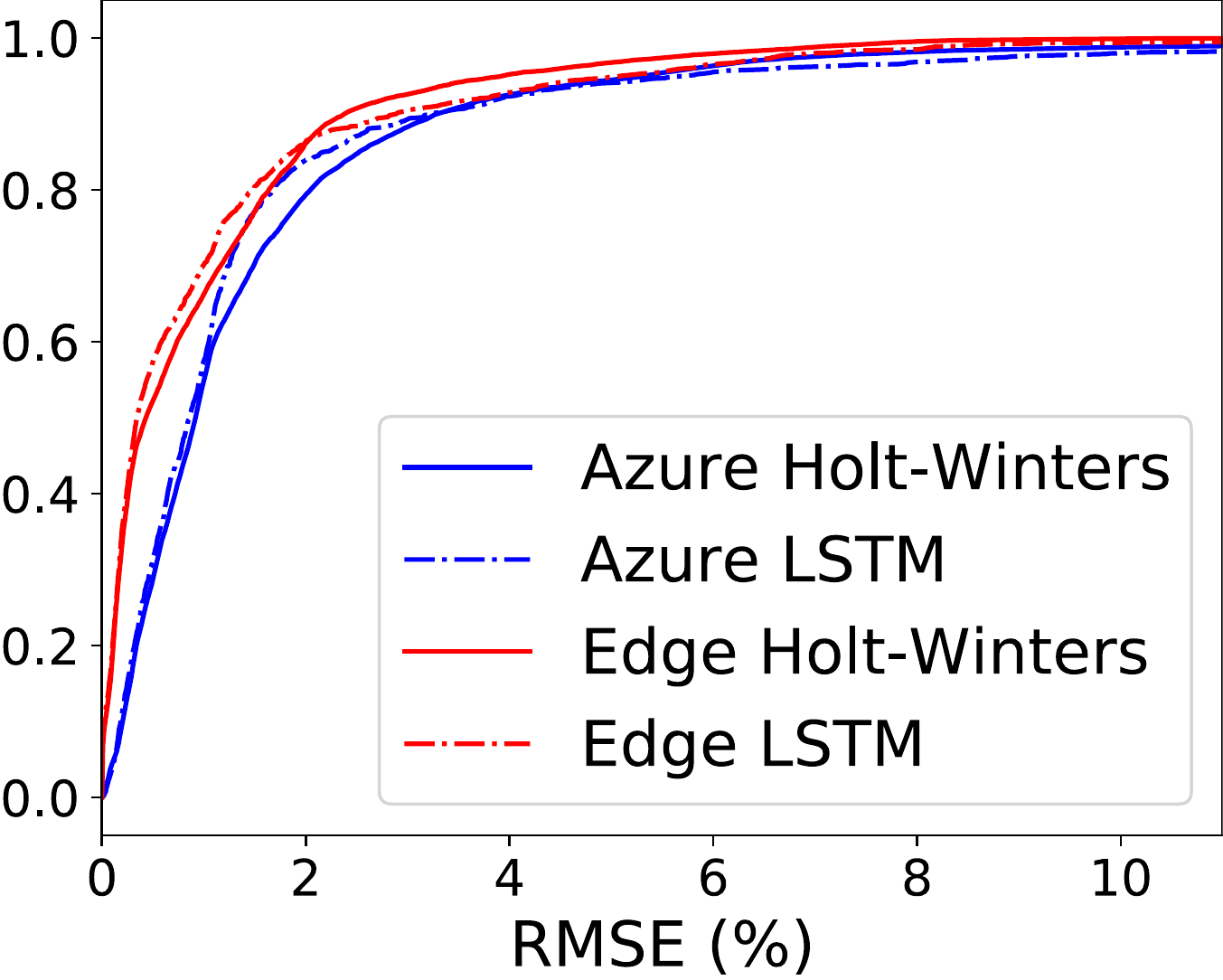}
		\subcaption{Mean CPU prediction}
		\label{fig:pred-usage-avg}
	\end{minipage}
	
	\caption{\textbf{Edge VMs' CPU usage is easier to predict than Cloud VMs.}
		Results accumulated across all VMs.
		Prediction time window length: half hour.
	}
	\vspace{-10pt}
	\label{fig:pred-usage}
\end{figure}

%% file: cost.tex
\subsection{Monetary Cost of Edge Apps}\label{sec:cost}

\input{fig-cost}

In this subsection, we investigate the monetary cost billed to edge customers for deploying edge apps on \sys.
For comparison, we also estimate the cost for the same hardware subscribed and workloads incurred on clouds.

\textbf{Baseline}
We use ``virtual baselines'' that simulate the situation if \sys's edge apps were deployed on cloud platforms.
It works by clustering and merging the VMs' usage (both hardware and bandwidth) of \sys into the \node distribution of cloud platforms based on geographical distances.
Here, we use two most popular cloud platforms in China: AliCloud (\textit{vCloud-1}) and Huawei Cloud (\textit{vCloud-2}).

\textbf{Billing model}
Appendix~\ref{appendix:billing} elaborates the detailed difference of their billing models.
To summarize, \sys and clouds charge the hardware resources (CPU, memory, storage) in a similar way.
For network, most cloud platforms support 3 kinds of billing models: by bandwidth (on-demand), by traffic quantity (on-demand), and by pre-reserved fixed bandwidth.
\sys currently only supports the first one, and even for this method, there are two notable differences.
\begin{itemize}[leftmargin=*]
\item \sys's network billing is much cheaper than \cloudsys in unit price, up to 13$\times$ depending on the geo-locations.
This is because edge servers handle requests from nearby locations, which means the traffic won't travel far along the network path.
This reduces the Internet backbone traffic, and leads to reduced operational cost for \sys and henceforth for the edge customers, compared to cloud platforms. 
\item \sys charges by the peak bandwidth usage per day, while \cloudsys charges in a more fine-grained way, i.e., peak bandwidth usage per minute.
This is in line with the billing model of \sys's ISP, likely because the ISP wants to mitigate the burstiness of the edge traffic that often exhibits high variance over time as shown in $\S$\ref{sec:vm-usage}.
\end{itemize}
To put it simply, \sys charges network usage in a cheaper yet less elastic way than clouds due to its traffic characteristics.


\textbf{Overview}
Table~\ref{tab:cost} summarizes the cost difference if the edge apps were moved from \sys to the cloud platforms (using the \sys's cost as the baseline).
Among the three network billing models, on-demand by bandwidth often costs less.
However, even compared to this model, \sys can significantly reduce the cost.
The average cost saving against vCloud-1/vCloud-2 is 45\% ($=1-1/1.82$)/43\% and up to 85\%/84\%.
This is because \sys has a cheaper bandwidth unit price than \cloudsys.
Only a few apps \sys charges more than \cloudsys.
Diving deeper, we find those apps either have high hardware resource demand or high bandwidth variance across time.
For example, an online education app has most of its traffic from 9:00 AM-12:00 PM.
Its peak (max) bandwidth usage is more than 10$\times$ higher than its average usage, while for other apps the variance is mostly between 1.5$\times$ and 4$\times$.
For those education apps, \cloudsys is more cost-friendly as it charges by minute while \sys charges by the peak bandwidth usage per day as aforementioned.

\textbf{Breakdown}
We then break down the bill to hardware and network bandwidth cost.
\sys often charges slightly more than \cloudsys (3\%--20\%) for each app's hardware resources, because of the relatively higher hardware maintaining cost on \sys currently.
However, \sys can significantly reduce the network bandwidth cost (up to 90\%), and the network resource often dominates the cost, i.e., 76\% on average and up to 96\%.
\revise{
Overall, the current customers of \sys are mostly video-related (as discussed in $\S$\ref{sec:vm-type}), of which the bandwidth cost is much higher than hardware resources.
In fact, cost efficiency is one of the major incentives for those customers to move their services from clouds to \sys.
}

\implications{
For \sys customers, deploying apps on its servers can significantly reduce the monetary cost due to cheaper bandwidth cost.
However, two kinds of apps may be exceptions: (1) apps with high hardware resource demand but less network demand; (2) apps with very high network usage variance across time.
For edge providers, it remains challenging to offer good billing elasticity to customers because of the high traffic variance across time.
}

%% file: fig-cost.tex
\begin{table}[t]
	\scriptsize
	\begin{tabular}{|l|l|r|r|r|}
		\hline
		\multicolumn{2}{|l|}{\textbf{\begin{tabular}[c]{@{}l@{}}Baselines normalized\\ to \sys (in times $\times$)\end{tabular}}} & \multicolumn{1}{l|}{\textit{\begin{tabular}[c]{@{}l@{}}On-demand,\\ by bandwidth\end{tabular}}} & \multicolumn{1}{l|}{\textit{\begin{tabular}[c]{@{}l@{}}On-demand,\\ by quantity\end{tabular}}} & \multicolumn{1}{l|}{\textit{\begin{tabular}[c]{@{}l@{}}Pre-reserved\\ (fixed)\end{tabular}}} \\ \hline
		\textbf{vCloud-1}                      & \begin{tabular}[c]{@{}l@{}}Range:\\ Mean:\\ Median:\end{tabular}                     & \begin{tabular}[c]{@{}r@{}}0.50$\times$--6.88$\times$\\ 1.82$\times$\\ 1.21$\times$\end{tabular}                        & \begin{tabular}[c]{@{}r@{}}0.60$\times$--14.98$\times$\\ 2.76$\times$\\ 1.97$\times$\end{tabular}                      & \begin{tabular}[c]{@{}r@{}}1.03$\times$--41.02$\times$\\ 4.93$\times$\\ 3.84$\times$\end{tabular}                         \\ \hline
		\textbf{vCloud-2}                      & \begin{tabular}[c]{@{}l@{}}Range:\\ Mean:\\ Median:\end{tabular}                     & \begin{tabular}[c]{@{}r@{}}0.64$\times$--6.43$\times$\\ 1.76$\times$\\ 1.25$\times$\end{tabular}                        & \begin{tabular}[c]{@{}r@{}}0.60$\times$--14.97$\times$\\ 2.66$\times$\\ 1.97$\times$\end{tabular}                      & \begin{tabular}[c]{@{}r@{}}1.03$\times$--14.87$\times$\\ 4.82$\times$\\ 3.56$\times$\end{tabular}                         \\ \hline
	\end{tabular}
\caption{\sys can significantly reduce the monetary cost compared to two cloud counterparts.
The cost includes both hardware and bandwidth.
The numbers are summarized over 50 heaviest apps.
}
\vspace{-10pt}
\label{tab:cost}
\end{table}

%% file: discuss.tex
\section{Implications}
\revise{
We summarize the key implications from our measurements.
Note that some of our recommendations to improve \sys may be common practices in cloud datacenters.
Yet, they remain an open problem in large-scale geo-distributed edge platforms as we have confirmed with the \sys development team.
}

\revise{
	\noindent \textbf{Killer apps for \sys-like edges}
	In the past decade, the research of edge computing is far ahead of its commercialization~\cite{mohan2020pruning}, mixed with real demands and hype.
	Looking into \sys, we find reducing network cost is, for now, a key incentive to move applications from clouds to edge platforms like \sys, making \textit{video-related applications} major customers, e.g., live streaming ($\S$\ref{sec:vm-type}).
	Other network-level metrics such as network latency and throughput also get improved with \sys to some extent.
	The improvement delivered to application-level QoE, however, can be diminished by many other practical factors beyond the network as demonstrated in $\S$\ref{sec:app-performance}.
	To avoid this, the hardware/software stacks of the edge infrastructure also need to be enhanced. This will allow \sys-like edges to better serve emerging computation-heavy applications such as AR/VR and autonomous driving.
}

\noindent \textbf{Sites as an integrated cluster}
Unlike cloud platforms, edge platforms have very limited per-site resources but a high density of site deployment, which necessitates cross-site coordination.
Such a need has been recently recognized by the clouds as well~\cite{tang2020twine}.
Treating edge \nodes as an integrated cluster facilitates the overall infrastructure management, but also brings unique challenges due to its nature of geo-distribution and ultra-low delay requirement.
Many of our observations (e.g., in $\S$\ref{sec:vm-usage} and $\S$\ref{sec:resource-allocate}) motivate cross-\node VM migration for balancing the resource usage and reducing resource fragmentation, etc.
While the technique has been extensively explored on both cloud~\cite{xvmotion,clark2005live,hines2009post} and edge~\cite{osanaiye2017cloud,rodrigues2016hybrid}, it remains challenging because
of the high migration delay and the impacts on the app QoS~\cite{akoush2010predicting,breitgand2010cost}. 

\noindent \textbf{Decomposing edge services}
While heavy IaaS VMs dominate the current usage of \sys, we believe the future of public edge platforms should embrace more elastic computing paradigms, e.g., microservices~\cite{nadareishvili2016microservice} and serverless computing~\cite{jonas2019cloud,aws-serverless}.
They help facilitate flexible resource management and fine-grained billing, which can benefit \sys as highlighted in $\S$\ref{sec:resource-allocate} and $\S$\ref{sec:cost}. 
However, such elasticity comes at a price.
For example, serverless computing has been criticized for its slow cold start~\cite{wang2018peeking,serverless-wild}.
Existing solutions to mitigate such slow start, e.g., highly optimized function loader and executor~\cite{du2020catalyzer,oakes2018sock,akkus2018sand}, can barely meet the requirements for ultra-low-delay edge applications.

\noindent \textbf{Cross-\nodes traffic scheduling}
Given massive, decentralized \nodes, which one should be responsible to handle a certain request from users?
Such a traffic scheduling strategy should not only satisfy the QoE of each application, but also consider cross-\node load balance.
The current scheduling policy of \sys, which is demonstrated to be oftentimes ineffective in $\S$\ref{sec:resource-allocate}, is owned by the developers (as clouds typically do).
On the opposite side, if the scheduling is done by the platform, it's difficult to guarantee the application QoE due to a lack of application-specific information.
We believe this is a new and open problem faced by edges that differ from clouds in many aspects, including the application programming model, resource allocation policy, and business model.

\revise{
\noindent \textbf{Workloads profiling and prediction} have been extensively studied for cloud platforms~\cite{cortez2017resource,gong2010press,calheiros2014workload}.
Directly applying them to \sys may not suffice considering the distinct workloads running atop them.
As a concrete example, scheduling VMs to different \nodes based on the \textit{past} resource usage prediction will likely lead to the \textit{future} usage change of that VM itself, as the usage pattern of edge VM highly depends on the geo-locations.
Such change may lead the prediction to be invalid shortly.
}

\section{Limitations and future work}\label{sec:discuss}

\noindent \textbf{Dataset representativeness}
Despite the best efforts we committed, the cloud-side workload dataset (Azure) still doesn't perfectly align with our \sys dataset, i.e., they were collected from different countries and at different times.
Thus, we tend to draw our conclusions conservatively.
Even so, we believe that the lessons learned from the study are valid for two reasons.
First, Azure is a global CSP with \nodes also deployed in China as \sys does.
Second, the Azure cloud workloads are relatively stable from its 2017 version to 2019 version; we therefore expect the workloads to exhibit similar characteristics in 2020 when the \sys dataset was collected.

\revise{
\noindent \textbf{Experimental settings}
Apart from the workloads traces, our actively collected dataset was collected by us through crowdsourcing-based controlled experiments.
We identify the following imperfections in carrying out those experiments.
\begin{myitemize}
\item In $\S$\ref{sec:network-delay}, we use ICMP-based ping instead of TCP to measure the network delay mainly because it's a built-in feature of non-rooted Android devices.
However, TCP-based ping is regarded more representative of normal workloads as ICMP is often treated with different priority by cloud providers than regular TCP/UDP traffic~\cite{hu2014need}.

\item In $\S$\ref{sec:app-performance} we implement two edge applications following their typical design and using state-of-art supporting libraries. 
Despite that, we admit that our (best-effort) implementation and deployment may not perfectly reflect those of commercial edge apps.
In the future, we will benchmark more applications and their diverse deployment configurations to comprehensively study how edge computing boosts the application QoE.

\item The number of endpoints (users) participating in our user study is limited. We plan to further scale it up in future work.

\item We did not investigate the network-layer characteristics (e.g., routing) of NEP traffic. Understanding them can facilitate network-edge cooperation through, for example, improved traffic engineering.

\end{myitemize}
}

\noindent \textbf{\sys as an early adopter}
\sys is now at an early stage (3 years since release) and not all its characteristics match what researchers commonly envision for ``ideal'' edge computing, e.g., in regard to the deployment density, customer diversity, and the VM elasticity feature.
Nonetheless, \sys is a leading edge platform whose number of sites is about two orders of magnitudes larger than a typical cloud provider, with large-scale adoption by a wide spectrum of commercial applications. Our measurements already suggest striking differences between NEP and cloud platforms.
Also, as commercial edge computing has recently made its debut, our results provide an important “baseline” for studying how it evolves in the future.
Note that this limitation is shared by other studies of emerging technologies such as 5G~\cite{5g-measurement,5g-mmwave}.

%% file: related.tex
\section{Related Work}

\noindent \textbf{Commercial edge/cloud platforms}
While the concept of ``pushing computations and services closer to users'' is generally accepted by edge researchers and practitioners, it still remains an open problem on how to bring the edges to reality.
In this work, we focus on a public edge platform \sys, which is regarded as an extension of traditional clouds but more diversely geo-distributed.
Other major cloud providers are also building their multi-tenant edge platforms, e.g., AWS Local Zones~\cite{aws-local-zones} and Azure Edge Zone~\cite{azure-edge-zone}.
However, those platforms are at a very early stage compared to \sys (Table~\ref{tab:density}), and there are no comprehensive measurements on them yet.
Major data providers like Facebook~\cite{taiji-sigcomm2020,schlinker2017engineering,SchlinkerCCSK19} and Google~\cite{yap2017taking} have built edge infrastructure (CDN, PoP, etc) to deliver their contents to end users more efficiently.
While content delivering is one killer use case in edge, \sys is built beyond the need for that but as a more general-purpose, multi-tenant computing platform like existing cloud providers.

\noindent \textbf{Measurements of edge/cloud platforms}
(1) At network performance aspect,
the wide-area network (WAN) performance has been extensively studied from the viewpoint of cloud providers, including the network latency~\cite{haq2017measuring,spring2003causes}, throughput~\cite{deng2018mobility,li2019large}, and resource demand volatility~\cite{kilcioglu2017usage}.
A few recent studies~\cite{mohan2020pruning,jin2019zooming} specifically target geo-distributed datacenters but are still at the cloud level.
Partly inspired by those work, we are the first to quantify the network and application performance of a real edge platform that has much denser DC deployment than traditional cloud platforms (shown in Table~\ref{tab:density}).
(2) At workloads aspect,
we are not aware of any prior work characterizing the workloads on edge platforms.
Some work~\cite{reiss2012heterogeneity,di2012characterization} analyze the first-party, container-based workloads on cloud platforms, which are orthogonal to ours that targets multi-tenant, VM-based workloads of \sys.
The most related work is performed on Azure~\cite{cortez2017resource} cloud, which is directly compared in this work.
As a key observation, we find that the edge workloads are indeed different from cloud.
\revise{
\cite{corneo2021surrounded} also performs large-scale measurements on the network performance among end users (8,000 RIPE Atlas probes) and datacenters (189 in total from many cloud providers).
Their study on global cloud platforms is orthogonal to ours on a much denser, nationwide edge platform.
}

\noindent \textbf{Edge systems and applications} have been built to bridge the gap between low-end devices and far-away clouds.
The key use cases include smart homes and cities~\cite{cicirelli2017edge,taleb2017mobile,trimananda2018vigilia},
autonomous driving~\cite{liu2020comparison,liu2020equinox,liu2019edge-autonomous-driving},
video analytics for smartphones~\cite{deepdecision,liu2019edge}, surveillance cameras~\cite{lavea,filterforward,vigil}, and drones~\cite{wang18sec}.
Those scattered thoughts can be regarded as motivations to build \sys that relieves edge developers from deploying and maintaining the edge hardware, just as the way cloud computing helps developers in the last twenty years.
Beyond specific use cases, there have been system-level optimizations towards edge performance and security~\cite{ren2020fine,miao2017streambox,miao2019streambox,park2019streambox}.
Those techniques are orthogonal to \sys.

%% file: conclusion.tex
\section{Conclusions}
We have performed the first comprehensive measurement on a commercial, multi-tenant edge platform.
Our study quantitatively answers two key questions: what is the edge performance perceived by end users and what are the edge workloads experienced by the edge operator. 
Our findings reveal critical differences between cloud and edge platforms; they also lead to insightful implications for designing future edge platforms and edge-based applications.

%% file: appendix-billing.tex
\appendix

\input{tab-pricing}


\section{Pricing Model Comparison}\label{appendix:billing}

Table~\ref{tab:pricing-model} shows a detailed comparison among the billing models of \sys and 2 popular cloud platforms (Alibaba Cloud and Huawei Cloud), referred to vCloud-1 and vCloud-2 in $\S$\ref{sec:cost}, in detail.
We focus on monthly cost, as it's generally supported by both cloud and edge platforms.

In \sys, the network traffic of VMs located in the same \node will be combined and charged together.
The bandwidth charged by ENS is the 95-th percentile daily peak bandwidth of the month.
In other words, \sys will first record the peak bandwidth usage per day, and then use the 4th highest one from all the daily peak usage in this month to generate the bill.
As previously discussed, such a billing method is less elastic as compared to cloud platforms that charge bandwidth by its average usage per minute or even second. 

%% file: tab-pricing.tex

\begin{table*}[]
	\scriptsize
	\begin{tabular}{|l|l|l|l|l|l|l|}
		\hline
		\multirow{2}{*}{\textbf{Unit: RMB}}                                                        & \multirow{2}{*}{\textbf{CPU}}                                                                 & \multirow{2}{*}{\textbf{Memory}}                                                              & \multirow{2}{*}{\textbf{Storage (SSD)}} & \multicolumn{3}{l|}{\textbf{Network}}                                                                                                                                                                                                                                                                                                                                                                                                          \\ \cline{5-7} 
		&                                                                                               &                                                                                               &                                         & \textbf{Sub-category}                                             & \textbf{Method}                                                                                                                                                                                           & \textbf{Example}                                                                                                                                               \\ \hline
		\multirow{3}{*}{\textbf{\begin{tabular}[c]{@{}l@{}}Alibaba\\ Cloud\end{tabular}}} & \multicolumn{2}{l|}{\multirow{3}{*}{\begin{tabular}[c]{@{}l@{}}2CPU + 4GB: 187/month\\ 2CPU + 8GB: 240/month;\\ 2CPU + 16GB: 318/month;\\ etc..\end{tabular}}}                                & \multirow{3}{*}{1/GB/month}             & \begin{tabular}[c]{@{}l@{}}Pre-reserved\\ (fixed)\end{tabular}    & \begin{tabular}[c]{@{}l@{}}1Mbps: 23/Mbps/month; \\ 2Mbps: 46/Mbps/month; \\ 3Mbps: 71/Mbps/month; \\ 4Mbps: 96/Mbps/month; \\ 5Mbps: 125/Mbps/month; \\ \textgreater{}5Mbps: 80/Mbps/month.\end{tabular} & \begin{tabular}[c]{@{}l@{}}2Mbps: 46/month;\\ 7Mbps: 125 + (7 - 5) * 80 = 285/month.\end{tabular}                                                              \\ \cline{5-7} 
		& \multicolumn{2}{l|}{}                                                                                                                                                                         &                                         & \begin{tabular}[c]{@{}l@{}}On-demand,\\ by bandwidth\end{tabular} & \begin{tabular}[c]{@{}l@{}}1$\sim$5Mbps: 0.063/Mbps/hour; \\ \textgreater{}5Mbps: 0.248/Mbps/hour.\end{tabular}                                                                                           & \begin{tabular}[c]{@{}l@{}}2Mbps: (24 * 30) * (2 * 0.063) = 90.72/month;\\ 7Mbps: (24 * 30) * {[}(2 * 0.063) + (7 - 5) * 0.248{]} = 447.84/month.\end{tabular} \\ \cline{5-7} 
		& \multicolumn{2}{l|}{}                                                                                                                                                                         &                                         & \begin{tabular}[c]{@{}l@{}}On-demand,\\ by quantity\end{tabular}  & 0.8/GB                                                                                                                                                                                                    & 1GB: 1 * 0.8 = 0.8                                                                                                                                             \\ \hline
		\multirow{3}{*}{\textbf{\begin{tabular}[c]{@{}l@{}}Huawei\\ Cloud\end{tabular}}}  & \multicolumn{2}{l|}{\multirow{3}{*}{\begin{tabular}[c]{@{}l@{}}1CPU + 1GB: 32.2/month;\\ 1CPU + 2GB: 72.2/month;\\ 2CPU + 4GB: 152.2/month;\\ 2CPU + 8GB: 251.6/month;\\ etc..\end{tabular}}} & \multirow{3}{*}{0.7/GB/month}           & \begin{tabular}[c]{@{}l@{}}Pre-reserved\\ (fixed)\end{tabular}    & \begin{tabular}[c]{@{}l@{}}1$\sim$5Mbps: 23/Mbps/month; \\ \textgreater{}5Mbps: 80/Mbps/month.\end{tabular}                                                                                         & \begin{tabular}[c]{@{}l@{}}2Mbps: 46/month;\\ 7Mbps: 23*5 + (7 - 5) * 80 = 275/month.\end{tabular}                                                             \\ \cline{5-7} 
		& \multicolumn{2}{l|}{}                                                                                                                                                                         &                                         & \begin{tabular}[c]{@{}l@{}}On-demand,\\ by bandwidth\end{tabular} & \begin{tabular}[c]{@{}l@{}}1$\sim$5Mbps: 0.063/Mbps/hour; \\ \textgreater{}5Mbps: 0.25/Mbps/hour.\end{tabular}                                                                                            & \begin{tabular}[c]{@{}l@{}}2Mbps: (24 * 30) * (2 * 0.063) = 90.72/month;\\ 7Mbps: (24 * 30) * {[}(5 * 0.063) + (7 - 5) * 0.25{]} = 586.8/month.\end{tabular}   \\ \cline{5-7} 
		& \multicolumn{2}{l|}{}                                                                                                                                                                         &                                         & \begin{tabular}[c]{@{}l@{}}On-demand,\\ by quantity\end{tabular}  & 0.8/GB                                                                                                                                                                                                    & 1GB: 1 * 0.8 = 0.8                                                                                                                                             \\ \hline
		\multirow{2}{*}{\textbf{\sys}}                                                    & \multirow{2}{*}{65/CPU/month}                                                                & \multirow{2}{*}{20/GB/month}                                                                  & \multirow{2}{*}{0.35/GB/month}          & \begin{tabular}[c]{@{}l@{}}Telecom or\\ Unicom\end{tabular}       & 25--50/Mbps/month                                                                                                                                                                                         & \begin{tabular}[c]{@{}l@{}}guangzhou-telecom 2Mbps: 50*2=100/month;\\ chengdu-telecom 2Mbps: 25*2=50/month.\end{tabular}                                       \\ \cline{5-7} 
		&                                                                                               &                                                                                               &                                         & CMCC                                                              & 15--30/Mbps/month                                                                                                                                                                                         & \begin{tabular}[c]{@{}l@{}}guangzhou-cmcc 2Mbps: 30*2=60/month;\\ chengdu-cmcc 2Mbps: 15*2=30/month.\end{tabular}                                              \\ \hline
	\end{tabular}

\caption{
A detailed comparison of the billing models of \sys and two popular cloud platforms in China.
The price of edge network bandwidth varies across different cities and operators.}
\label{tab:pricing-model}
\end{table*}